\documentclass[11pt,letterpaper]{article}
\usepackage[margin=1in]{geometry}
\usepackage[T1]{fontenc}
\usepackage{lmodern,microtype,amsmath,amssymb,amsthm,mathtools}
\usepackage{graphicx,booktabs,tabularx,array,enumitem}
\usepackage{xcolor}
\usepackage{tikz}
\usepackage{needspace}
\usepackage[hidelinks]{hyperref}
\usepackage[nameinlink,capitalize,noabbrev]{cleveref}
\hypersetup{pdftitle={Impatience Makes Matching NP-Hard}}
\setlist{itemsep=.2em,topsep=.35em,leftmargin=1.6em}
\renewcommand{\arraystretch}{1.08}

\allowdisplaybreaks
\newtheorem{theorem}{Theorem}[section]
\newtheorem{lemma}[theorem]{Lemma}
\newtheorem{proposition}[theorem]{Proposition}
\newtheorem{corollary}[theorem]{Corollary}
\theoremstyle{definition}

\theoremstyle{remark}

\newcommand{\E}{\mathbb E}
\newcommand{\Prob}{\mathbb P}

\newcommand{\ind}[1]{\mathbf{1}_{\{#1\}}}

\newif\ifreview
\reviewtrue
\definecolor{RevisionRed}{RGB}{180,35,42}
\definecolor{ReplyGreen}{RGB}{0,105,75}

\title{\LARGE\bfseries 
Impatience Makes Matching NP-Hard}
\author{%
  Chen Yan\\
  Department of Electrical and\\
  Computer Engineering\\
  The Ohio State University\\
  \texttt{yan.1964@osu.edu}
  \and
  Sushil Mahavir Varma\\
  Department of Industrial and\\
  Operations Engineering\\
  University of Michigan, Ann Arbor\\
  \texttt{sushilv@umich.edu}
}
\date{}
\date{}
\begin{document}
\maketitle

\begin{abstract}

\normalsize
Motivated by emerging applications such as online marketplaces and quantum switches,
we study a dynamic bipartite matching model with impatient agents, where the two sides represent customers and suppliers. Agents of each type arrive according to an independent Poisson process, and the platform decides which pairs of agents to match. The agents are possibly impatient and may leave the system if not matched before their patience expires. The platform aims to maximize throughput, i.e., the number of matches per unit time. In this paper, we reveal an interesting phase transition in the computational hardness of throughput maximization. In particular, 
we show that approximating the optimal long-run throughput
within a sufficiently small fixed relative error is NP-hard,
even when all agents have rate-one exponential patience.
This result implies a computational barrier to constructing near-optimal, efficiently executable matching policies that induce a matching process with fast mixing. On the other hand, when all agents on at least one side have infinite patience, we provide a simple polynomial-time algorithm that maximizes the throughput.
\end{abstract}

\clearpage

\section{Introduction}\label{sec:introduction}

Matching is a central operational problem in many markets and service
systems. Ride-hailing platforms match passengers with drivers, and
platforms operating autonomous vehicles must similarly assign requests
to available vehicles~\cite{WZZ24,GWDC24}. Online marketplaces match
workers with tasks~\cite{CILBN20,AS22,VBMW23}; organ-allocation systems
and kidney exchanges match patients with compatible
organs~\cite{Leshno22,Unver10,ALOG20}; and quantum switches combine
short-lived quantum resources to serve requests~\cite{ZJM26}.
Although these applications differ in their operational details, they
share a basic challenge: matching opportunities do not remain available
indefinitely. Waiting agents may leave, and resources may expire before
they can be used. These settings motivate a common abstraction in which
agents abandon if they are not matched before
their patience expires. The resulting algorithmic question is how to
match impatient agents to maximize reward earned per unit time, or throughput when each match earns unit reward.

In this paper, we study a bipartite matching model with customers on one side and
suppliers on the other. Each vertex represents an agent type and has
a queue of waiting agents; an edge indicates that its endpoint types
are compatible. Customers and suppliers arrive over time and join
their respective queues. The platform decides when to match agents and
which compatible pairs to select. Once matched, both agents leave the
system instantaneously. On a successful match, the platform receives an edge-dependent reward. An unmatched agent may instead abandon when
its patience expires. Our principal objective is the long-run average reward. When all the edge rewards are one, the long-run average reward specializes to throughput, i.e., the average number of matches per unit time.

\begin{figure}[htbp]
  \centering
  \begin{tikzpicture}[
    x=1cm,y=1cm,
    every node/.style={font=\normalsize},
    agent/.style={circle,draw,thick,minimum size=9mm,inner sep=1pt}
  ]
    \node at (0,2.25) {Suppliers};
    \node at (5,2.25) {Customers};

    \node[agent,fill=black!12] (s1) at (0,1.5) {$s_1$};
    \node[agent] (s2) at (0,0) {$s_2$};
    \node[agent] (c1) at (5,1.5) {$c_1$};
    \node[agent,fill=black!12] (c2) at (5,0) {$c_2$};

    \draw[thick] (s1) -- (c1);
    \draw[thick] (s2) -- (c2);
    \draw[thick,dashed] (s1) --
      node[midway,above,sloped,fill=white,inner sep=2pt]
      {match now} (c2);
  \end{tikzpicture}
  \caption{An $N$-network illustrating the reservation tradeoff.
  Shaded vertices indicate the two currently available agents.
  Matching them uses the dashed edge; waiting preserves two potential
  future matches along the solid edges, but exposes both agents
  to abandonment.}
  \label{fig:n-network}
\end{figure}
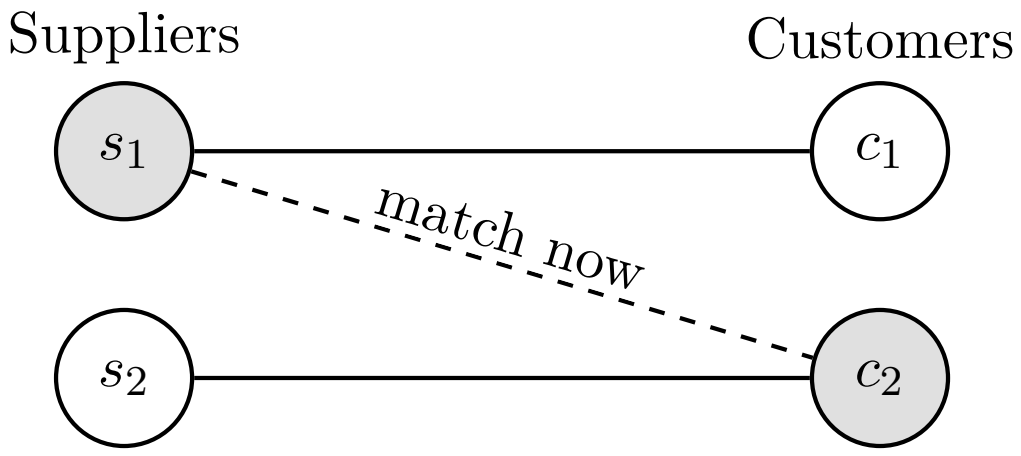

For this long-run objective, the problem without abandonment is
tractable: a static weighted-flow optimization determines the optimal
value, and a simple routing policy attains it. Does abandonment change
this computational picture? The $N$-network in
Figure~\ref{fig:n-network} illustrates the underlying tradeoff.
Suppose one supplier of type $s_1$ and one customer of type $c_2$ are
waiting. Matching them along the diagonal edge completes one match
immediately. Waiting may instead allow two matches along the horizontal
edges: the supplier can serve a future $c_1$ customer, while the customer
can use a future $s_2$ supplier. But if agents are impatient, either
waiting agent may abandon before these opportunities arise. The
platform must therefore weigh the value of preserving compatibility
for future arrivals against the risk of losing the agents it retains.

Despite substantial work on dynamic matching and matching-queue
control~\cite{GW14,NS19}, including approximation algorithms with
stochastic arrivals and departures~\cite{CILBN20,AS22}, the computational
complexity of the optimal value has remained unresolved in basic
impatient matching models. In particular,
\cite[Section~1.3]{AAAPS26} identifies the NP-hardness of stationary
bipartite matching, where suppliers have finite patience and customers must
be served immediately, as an open question. In this paper, we resolve this question
for the optimal numerical value and establish NP-hardness when both
sides have finite patience as well.

More precisely, we show that estimating the optimal throughput within a sufficiently small universal relative error $\delta>0$ is NP-hard, even with rate-one exponential patience on both sides. The same
value-hardness result holds with rate-one supplier patience and zero
customer patience. Consequently, neither regime admits a
polynomial-time approximation scheme (PTAS) for the numerical optimal
value unless $\mathrm P=\mathrm{NP}$.

These results lead to several hardness consequences for policy design on matching problems with abandonment.
For a sufficiently small universal constant $\delta>0$, unless
$\mathrm{NP}=\mathrm{RP}$, no polynomial-time procedure can construct,
with probability at least $2/3$ over policy construction, a stationary
Markov policy that simultaneously achieves throughput at least
$(1-\delta)$ times the optimal value, mixes in polynomial time
from the empty state, and admits efficient execution.
Thus, a PTAS producing efficiently executable stationary Markov
policies cannot guarantee both good performance and polynomial mixing.
Furthermore, for admissible policies that may be history-dependent,
a similar result shows that no polynomial-time procedure can construct,
with probability at least $2/3$ over policy construction, an efficiently
executable policy that achieves close-to-optimal throughput and whose
finite-horizon expected average reward approaches and remains close
to that throughput within polynomial time, unless
$\mathrm{NP}=\mathrm{RP}$.

\subsection{Main results and proof ideas}
We consider three patience categories on each side of the bipartite graph: infinite patience,
finite positive exponential patience, and zero patience. An infinitely
patient agent waits until matched; a zero-patience agent must be matched
on arrival or leave immediately. Every type on a side has the indicated
category. The nine combinations, or six regimes up to interchanging
the sides, are summarized in Table~\ref{tab:classification}.

\begin{table}[htbp]
  \centering
  \renewcommand{\arraystretch}{1.16}
  \setlength{\tabcolsep}{7pt}
  \begin{tabular}{@{}lccc@{}}
    \toprule
    & \multicolumn{3}{c}{Customer patience} \\
    \cmidrule(l){2-4}
    Supplier patience & Infinite & Finite positive & Zero \\
    \midrule
    Infinite        & Polynomial & Polynomial & Polynomial \\
    Finite positive & Polynomial & NP-hard    & NP-hard \\
    Zero            & Polynomial & NP-hard    & Trivial \\
    \bottomrule
  \end{tabular}
  \caption{Complexity of computing the optimal long-run average
reward. ``Polynomial'' means that the numerical optimum is
computable in polynomial time. In the hard regimes,
approximation within a sufficiently small fixed relative error
is NP-hard.}
  \label{tab:classification}
\end{table}

The source of hardness is easiest to see when customers have zero
patience and suppliers have finite positive patience. A customer is
lost immediately if no compatible supplier is available on arrival.
To protect future demand, the platform therefore wants to maintain
supplier inventory covering the different customer types. But each
supplier held in reserve may abandon, so retaining more inventory also
wastes supply. The platform must balance uncovered demand against
abandonment of the suppliers it keeps. Our reductions connect the
problem of providing coverage with few occupied supplier types to the combinatorial problem of finding a minimum vertex cover of a graph. 

\begin{figure}[htbp]
  \centering
  \includegraphics[width=0.95\linewidth]{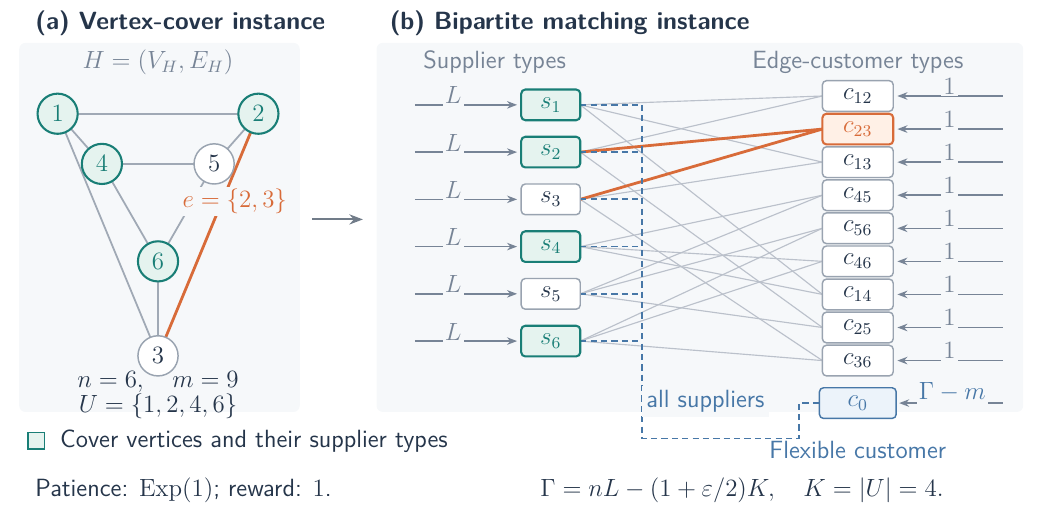}
  \caption{The Finite-Finite construction in
  Theorem~\ref{thm:ff-hardness}. Each source vertex becomes a supplier
  type, each source edge becomes a customer type compatible with its
  endpoints, and the flexible customer type can use any supplier.
  The highlighted cover $U=\{1,2,4,6\}$ specifies where a good policy should
  reserve some available suppliers for edge customers.}
  \label{fig:reduction}
\end{figure}

Start with a simple cubic graph $H$, where each vertex has degree $3$. As illustrated in
Figure~\ref{fig:reduction}, create a supplier type for every vertex and
an edge-customer type for every edge. A customer associated with an
edge can match only with suppliers at its two endpoints. Supplier types
arrive at a common rate $L$, edge-customer types arrive at rate one,
and an additional flexible customer type can match with every supplier
type. 

In this construction, having a compatible supplier available for
every edge-customer type is equivalent to the occupied supplier
vertices forming a vertex cover of $H$. Finding the fewest suppliers
needed for this coverage therefore amounts to finding a
minimum vertex cover. 
A small cover permits a policy that protects demand while
reserving fewer suppliers; a large minimum vertex-cover size
forces losses under every admissible policy.
By choosing the flexible customer's arrival rate appropriately, we transfer the hard problem of approximating a minimum vertex cover~\cite{Chlebik2006} to the problem of approximating
optimal matching throughput.

Hence, on a source instance with a small vertex cover, sufficiently high finite-horizon
reward implies that the occupied supplier vertices at a
uniformly chosen time can be completed to a small vertex cover
with constant probability. If a suitable policy can be
constructed and executed, and reaches this reward level within
polynomial time, the observation gives a one-sided randomized
polynomial-time algorithm for the source promise,
leading to $\mathrm{NP}=\mathrm{RP}$. For
stationary Markov policies, a polynomial mixing bound at a
sufficiently small fixed accuracy yields the same consequence.

In contrast to these negative results, if at least one side has infinite patience, an optimal weighted deterministic flow
gives a routing policy with equal-rate arrival streams on each edge.
An infinitely patient endpoint stores agents for that edge, and arrivals
at the other endpoint match whenever its inventory is nonempty. Thus, the dynamics on each edge resemble a single-server queue \textit{without} abandonment: arrivals of impatient agents correspond to service opportunities, and arrivals of patient agents correspond to customer arrivals in the single-server queue.
The expected reward loss through empty inventories is $O(\sqrt{T})$ by
time $T$, so the policy attains the optimal long-run reward. Both the
flow and a policy description are computable in polynomial time.

Finally, if both sides have zero patience, no matches occur under our
arrival model, completing the classification in Table~\ref{tab:classification}.

\subsection{Related work}

\paragraph{Dynamic matching and control.}
Our work builds on the study of dynamic matching and matching-queue
control~\cite{GW14,NS19}. In the absence of abandonment, greedy policies
can have bounded hindsight regret under a general-position gap
condition~\cite{KAG25}. These results address how to make effective
matching decisions over time when the agents are patient. 
On the other hand, with abandonment,
\cite{CILBN20,AS22} develop approximation algorithms for dynamic
weighted matching with heterogeneous arrival and departure parameters.

\paragraph{Approximation guarantees and their benchmarks.}
In a general dynamic matching model, \cite{ANNS26} gives a
polynomial-time static priority policy attaining at least half the
reward of an omniscient policy, and shows that this ratio is tight.  
The omniscient optimum upper bounds the optimal online reward,
so this also gives a $1/2$-approximation to optimal online
reward. 

In the setting corresponding to our Finite-Zero regime, where customers have zero patience and suppliers have finite patience, \cite{AAAPS26} gives a
$(1-1/e+\delta_0)$-approximation to optimal online reward for a
universal $\delta_0>0$. 
Stronger approximations are possible in structured subclasses.
Notably, \cite{AAS25} study matching costs subject to an
average-match-rate constraint, with common supplier abandonment.
They provide a bicriteria FPTAS on general compatibility graphs when the number of supplier queues is fixed. 
Our hardness results complement these guarantees: when the
number of supplier types is part of the input, approximating
the optimal numerical throughput within a sufficiently small
fixed relative error is NP-hard.

\begin{table}[htbp]
  \centering
  \renewcommand{\arraystretch}{1.16}
  \begin{tabularx}{\linewidth}{
    @{}X>{\centering\arraybackslash}p{0.24\linewidth}l@{}
  }
    \toprule
    Setting & Fraction of optimal online reward & Reference \\
    \midrule
    General matching; all types have finite positive patience
      & $1/2$ & \cite{ANNS26} \\
    Bipartite Finite-Zero matching
      & $1-1/e+\delta_0$ & \cite{AAAPS26} \\
    Finite-Zero; fixed constant number of supplier types
      & $1-\varepsilon$ & \cite{AAS25} \\
    \bottomrule
  \end{tabularx}
  \caption{Polynomial-time policy guarantees relative to the optimal
nonanticipating policy. In the second row $\delta_0>0$ is universal.
The last row assumes unit rewards and a common supplier
abandonment rate.}
  \label{tab:online-ratios}
\end{table}

\paragraph{Patience, inventory, and stationary behavior.}
The effect of impatience is studied quantitatively in~\cite{KG25},
which develops nonasymptotic scaling laws and identifies four
operating regimes for a two-type matching queue. With only one
possible match type, that model isolates inventory fluctuations and
abandonment without the allocation decisions induced by a growing
compatibility graph. Our classification instead concerns how patience
changes the computational difficulty of those decisions. Related
queueing work studies stability under paired discrete-time
arrivals~\cite{BGM13} and obtains product-form stationary distributions
for FCFS matching in an $N$-system with reneging~\cite{CNY20}.
Those analyses characterize stability or performance under specified
rules, rather than the computational complexity of optimizing over
all admissible policies. This distinction also matters in our
infinite-patience regimes, where a policy can attain optimal average
reward \emph{without} a stationary inventory distribution.

\paragraph{Large-market limits and resource matching.}
Asymptotic approaches provide tractability in other directions.
Fluid models and asymptotically optimal policies for large markets
with heterogeneous, impatient supply and demand are developed
in~\cite{ADSW25}, allowing more general patience distributions.
A heavy-traffic Brownian approximation for a discounted objective
with matching rewards, holding costs, and abandonment penalties
is studied in~\cite{AX25}. Such results concern performance under
scaling limits.
Quantum-switch
models~\cite{ZJM26,Srikant26} and related decision-dependent networks~\cite{BVW25} offer another related setting,
where resources are perishable and a service operation may consume
several resources together with a request. These models motivate the
role of limited availability, but their capacity and stability
questions, and potentially multi-way matches, differ from our
bipartite matching problem.

\subsection{Terminology and notation}
Running time is measured in the binary input length. The class $\mathrm P$ consists of decision problems with a YES/NO answer solvable
in deterministic polynomial time.
The class $\mathrm{NP}$ consists of decision problems whose
YES inputs have polynomial-length certificates verifiable
in deterministic polynomial time, whereas NO inputs have
no accepted certificates.
The class $\mathrm{RP}$ consists of decision problems solvable
by a randomized algorithm in polynomial worst-case time that
never accepts a NO input and accepts every YES input with
probability at least $2/3$.
The class $\mathrm{PSPACE}$ consists of decision problems
solvable by a deterministic algorithm using polynomial space.
The standard inclusions are
\[
 \mathrm P\subseteq\mathrm{RP}\subseteq\mathrm{NP}
 \subseteq\mathrm{PSPACE}.
\]
The first two hold because deterministic algorithms need no
random bits, and an accepting sequence of random bits provides
an NP certificate. The last holds because polynomial-length
certificates can be enumerated and checked one at a time using
polynomial space.

A problem is NP-hard if every problem in $\mathrm{NP}$ admits
a deterministic polynomial-time many-one reduction to it;
it is NP-complete if it also belongs to $\mathrm{NP}$.
Likewise, a problem is PSPACE-hard if every problem in
$\mathrm{PSPACE}$ admits such a reduction to it;
it is PSPACE-complete if it also belongs to $\mathrm{PSPACE}$.

We write $\mathcal L(X)$ for the probability law of a random
variable $X$. For probability measures $\mu$ and $\nu$ on the
same measurable space, their total-variation distance is
$\|\mu-\nu\|_{\mathrm{TV}}:=\sup_A|\mu(A)-\nu(A)|$,
where $A$ ranges over measurable sets.
A vertex cover of a graph is a set of vertices containing at least one endpoint of every edge.
Maximizing throughput is equivalent to maximizing the average
reward when each match earns one unit of reward.

\section{The Matching Model with Impatient Agents} \label{sec:model}

\subsection{Arrivals, compatibility, and patience}

Let $G=(S,C,E_G)$ be a finite bipartite compatibility graph,
where $S$ and $C$ are the supplier and customer types.
Suppliers of type $i\in S$ arrive according to a Poisson process
of rate $\lambda_i>0$, and customers of type $j\in C$ arrive
according to a Poisson process of rate $\gamma_j>0$.
All arrival processes are independent.
The platform may match any compatible pair of agents currently
in the system. A match between types $i$ and $j$ earns a
nonnegative reward $w_{ij}$, and both matched agents leave
immediately.

Let $\mu_i^s,\mu_j^c\in[0,\infty]$ denote the patience
parameters of supplier type $i$ and customer type $j$.
A finite positive parameter is the rate of an exponential
patience time. A parameter of zero means infinite patience:
the agent remains until matched. A parameter of infinity means
zero patience: the agent must be matched on arrival or leave
immediately. All patience times are independent of one another
and of the arrival processes, and their realizations are hidden
from the controller.

For our hardness results, we consider the following two regimes:
\begin{itemize}
    \item \textbf{Finite-Finite:} All supplier and customer types abandon with rate $1$.
    \item \textbf{Finite-Zero:} All suppliers abandon with rate $1$ and all customers abandon if not matched as soon as they arrive.
\end{itemize}
For our positive results, we consider either $\mu^s_i = 0$ or $\mu^c_j = 0$ for all $(i, j) \in E_G$, i.e., every edge is incident to an infinitely patient agent. This setting captures the following three regimes of Table~\ref{tab:classification}:
\begin{itemize}
    \item \textbf{Infinite-Infinite}: All agents are infinitely patient, so no one leaves the system until matched.
    \item \textbf{Infinite-Finite}: All suppliers stay in the system until matched and the customers abandon at finite rates.
    \item \textbf{Infinite-Zero}: All suppliers stay in the system until matched and the customers abandon if not matched as soon as they arrive.
\end{itemize} Lastly, note that the case of \textbf{Zero-Zero} is trivial, i.e., when agents on both sides abandon if not matched immediately, the platform makes no matches with probability 1 because no compatible supplier-customer pair will arrive at the same time.

The graph, arrival rates, rewards, and finite positive
abandonment rates are part of the input. All finite rates and
rewards are rational, and the patience categories are explicitly
specified. We also define
\[
 \Lambda:=\sum_{i\in S}\lambda_i+\sum_{j\in C}\gamma_j
\]
for the total arrival rate. The system starts empty. We denote by $Q_i^s(t)$ the number of suppliers of type $i$ waiting in the system at time $t \in \mathbb{R}_+$, i.e., the queue length of supplier type $i \in S$. Similarly, $Q_j^c(t)$ denotes the queue length of customer type $j \in C$ at time $t \in \mathbb{R}_+$. 
Let
\[
 Q(t):=\bigl((Q_i^s(t))_{i\in S},
            (Q_j^c(t))_{j\in C}\bigr)
 \in\mathbb Z_+^{|S|+|C|}
\]
be the post-decision queue-length vector.

\subsection{Policies and the reward objective}

An admissible policy is nonanticipating: its decisions at each
arrival or abandonment depend only on the observed history up to
that event and on internal randomization independent of the arrival
processes and patience times. At each such event, the policy may
make a finite sequence of instantaneous matches. It may also
decline a currently available match, so strategic idling is
permitted. We allow queues to accumulate unbounded inventory;
stability is not required.

Let $M_{ij}^\pi(T)$ count matches on edge $(i,j)$ by time $T$ under policy
$\pi$. The total match count and total reward are, respectively,
\[
 M^\pi(T):=\sum_{(i,j)\in E_G}M_{ij}^\pi(T),\qquad
 W^\pi(T):=\sum_{(i,j)\in E_G}w_{ij}M_{ij}^\pi(T).
\]
Define
\begin{equation}
 r_\pi(T):=\frac{\E W^\pi(T)}{T},\qquad
 g^\pi:=\liminf_{T\to\infty}r_\pi(T),\qquad
 g^*:=\sup_\pi g^\pi.
 \label{eq:gain}
\end{equation}
As every match uses two distinct arrivals,
we get $0\le g^*\le w_{\max}\Lambda/2$, where
$w_{\max}:=\max\{w_{ij}:(i,j)\in E_G\}$. For our hardness results, we shall set all rewards to one, i.e., $w_{ij} = 1$ for all $(i, j) \in E_G$, and so
$W^\pi=M^\pi$ and $g^*\le\Lambda/2$. In this case, we shall call $g^*$ the optimal throughput.

Maximizing the average reward can be viewed as an
infinite-horizon average-reward Markov decision process
(MDP) with state $Q(t)$~\cite{GHL09}.
A stationary Markov policy chooses matches using the current
queue lengths, the arrival or abandonment event just observed,
and fresh randomization. Under such a policy, the queue-length
process $\{Q(t):t\ge0\}$ is a continuous-time Markov chain (CTMC).


\section{Main Results}
In this section, we present the main results of the paper concerning a polynomial-time algorithm for the patient regimes and hardness results for the impatient regimes.

\subsection{Infinite patience on a vertex cover}
Suppose every compatibility edge has an infinitely patient endpoint. Consider the following weighted flow problem:
\begin{equation}\label{eq:flow-lp}
 F^*=\max\left\{\sum_{(i,j)\in E_G}w_{ij}f_{ij}:
 \begin{array}{ll}
 f_{ij}\ge0,&(i,j)\in E_G,\\
 \sum_{j:(i,j)\in E_G}f_{ij}\le\lambda_i,&i\in S,\\
 \sum_{i:(i,j)\in E_G}f_{ij}\le\gamma_j,&j\in C
 \end{array}\right\}.
\end{equation}
The variable $f_{ij}$ represents a long-run matching rate on
edge $(i,j)$. The constraints bound the total matching rate at
each vertex by its arrival rate. Expected match counts divided
by $T$ satisfy these constraints~\cite{GW14,KAG25}, so
$r_\pi(T)\le F^*$ for every admissible policy and every horizon.
The following theorem establishes attainability of this bound.

\begin{theorem}[Infinite patience on a vertex cover] \label{thm:permanent}
Suppose that every edge of $E_G$ has an infinitely patient endpoint.
Then $g^*=F^*$. Furthermore, there exists a policy $\pi$ that can be described and computed in polynomial time and that attains $g^*$ exactly. More specifically, if $f$ is an optimal flow for \eqref{eq:flow-lp}, policy $\pi$ satisfies
\begin{equation}\label{pos:eq:rate-bound}
 0\le F^*-r_\pi(T)
 \le2\sqrt{\frac{2}{T}}
       \sum_{(i,j)\in E_G}w_{ij}\sqrt{f_{ij}}
 \qquad(T>0).
\end{equation}
\end{theorem}

An entire side of a bipartite graph is a vertex cover, so the
theorem gives the Infinite-Infinite, Infinite-Finite, and
Infinite-Zero cases in Table~\ref{tab:classification}.

The policy is described in Appendix~\ref{app:positive}.
At each supplier arrival of type $i$, it chooses edge $(i,j)$
with probability $f_{ij}/\lambda_i$; at each customer arrival
of type $j$, it chooses edge $(i,j)$ with probability
$f_{ij}/\gamma_j$. Each edge has an infinitely patient endpoint
that stores agents for matches along that edge.
The two selected arrival streams have the same rate.
The finite-horizon bound in \eqref{pos:eq:rate-bound} shows that
the expected average reward converges to $F^*=g^*$.

\subsection{NP-Hardness of throughput optimization and its consequences}  \label{subsec:NP-hardness}
We now show that the matching problem \eqref{eq:gain} becomes NP-hard when both sides abandon, even with unit rewards and unit abandonment rates. In particular, set $w_{ij} \equiv 1$ for all $(i, j) \in E_G$ and consider the Finite-Finite or Finite-Zero setting as described in Section~\ref{sec:model}.

\begin{theorem}
\label{thm:np_hardness}
For each of the Finite-Finite and Finite-Zero regimes, there
exists a universal constant $\delta>0$ such that computing
a numerical estimate $\widehat g$ satisfying
\[
 |\widehat g-g^*|\le\delta g^*
\]
is NP-hard. Consequently, unless $\mathrm P=\mathrm{NP}$,
neither regime admits a polynomial-time approximation scheme
for the numerical optimal value.
\end{theorem}

This result establishes the hardness of approximately computing
the optimal throughput. For the Finite-Finite regime, the same
reduction also yields two consequences for policy design:
one for stationary Markov policies and the other for
history-dependent admissible policies.

To that end, we must account for the computational work needed
both to construct a policy and to run it: a policy can be easy to
describe and construct yet take exponential time to execute.
We say that a policy is \emph{efficiently executable} if it is
represented by a randomized program that, at each arrival or
abandonment, receives only the event type and current queue lengths,
chooses matches, and may retain information from past events.
Initialization and processing the first $j$ events must require
worst-case time and memory polynomial in $j$ and the combined
binary length of the instance and the program. One fixed
polynomial bound must apply to all policy programs under
consideration.

For the first result, note that, due to abandonment, every
stationary Markov policy in the Finite-Finite and Finite-Zero
settings induces a CTMC $\{Q(t):t\ge0\}$ that is positive recurrent
on the class of states reachable from empty queues.
For completeness, the details are presented in
Lemma~\ref{comp:recurrence} of
Appendix~\ref{app:technical-details}.
For such a policy $\pi$, denote its unique invariant distribution
on this class by $\nu_\pi$.
For $0<\rho<1$, define its mixing time from empty by
\begin{equation}
 t_{\mathrm{mix}}^{0,\pi}(\rho)
 :=\inf\left\{t\ge0:
   \sup_{s\ge t}
   \bigl\|\mathcal L(Q^\pi(s))-\nu_\pi\bigr\|_{\mathrm{TV}}
   \le\rho\right\}.
 \label{mix:definition}
\end{equation}
This quantity measures how soon the queue-length distribution,
starting empty, comes within total-variation distance $\rho$
of stationarity and remains within that distance.

In the following two corollaries, $B$ denotes the instance
bit length. The proofs of these two corollaries are deferred to Section~\ref{sec:cor-proofs}.

\begin{corollary}[Policy construction and polynomial mixing] \label{cor:policy-hardness}
There exist universal rational constants $\eta_0>0$ and
$\rho\in(0,1)$ with the following property.
Suppose a randomized polynomial-time algorithm returns an
admissible, efficiently executable stationary Markov policy
$\pi$ for every Finite-Finite reduction instance.
If, for some fixed polynomial $p$, its output satisfies
\begin{align*}
 g^\pi \ge g^*-\eta_0\Lambda, \qquad
 t_{\mathrm{mix}}^{0,\pi}(\rho) \le p(B)
\end{align*}
with probability at least $2/3$ over policy construction,
then $\mathrm{NP}=\mathrm{RP}$.
\end{corollary}

The second result replaces mixing by a bound on how quickly
expected average reward approaches its long-run value.
It allows the returned policy to depend on past events.

\begin{corollary}[Policy construction and finite-horizon performance]  \label{cor:deployment}
There exists a universal rational constant $\eta_0>0$ with
the following property.
Suppose a randomized polynomial-time algorithm returns an
admissible, efficiently executable policy $\pi$ for every
Finite-Finite reduction instance.
If, for some fixed polynomial $p$, its output satisfies
\[
 \begin{aligned}
  g^\pi &\ge g^*-\eta_0\Lambda,\\
  r_\pi(T) &\ge g^\pi-\eta_0\Lambda
       \qquad\text{for every }T\ge p(B)
 \end{aligned}
\]
with probability at least $2/3$ over policy construction,
then $\mathrm{NP}=\mathrm{RP}$.
\end{corollary}

\section{Proof details of NP-hardness for the Finite-Finite regime}  \label{sec:hardness}

In this section we prove Theorem~\ref{thm:np_hardness} for the Finite-Finite setting. The proof for the Finite-Zero setting is similar, and the details are presented in Appendix~\ref{app:Finite-Zero}. We focus on the Finite-Finite setting in the main body of the paper as it is more challenging and subsumes the main ideas used to resolve the Finite-Zero setting. The reduction starts from a vertex-cover problem on a graph $H$, which we call the \emph{source graph} to distinguish it from the
compatibility graph of the matching instance.

\subsection{The source problem and the matching construction}  \label{sec:construction}

\paragraph{The source gap and the compatibility graph.}

For a graph $H$, write $\tau(H)$ for its minimum vertex-cover size.
The bounded-degree vertex-cover gap is a standard source of approximation
hardness~\cite{PapadimitriouYannakakis1991}. We use the following consequence
of the cubic $100/99$ inapproximability theorem~\cite{Chlebik2006}.

\begin{theorem}[Cubic vertex-cover gap]\label{thm:cubic-source}
For the fixed constant $\varepsilon=1/100$, it is NP-hard to
distinguish
\begin{equation}
 \mathrm{YES}: \tau(H)\le K\qquad\text{from}\qquad
 \mathrm{NO}: \tau(H)\ge(1+\varepsilon)K
 \label{eq:source-promise}
\end{equation}
on finite simple cubic graphs with $n$ vertices and an integer
threshold $K$ satisfying $n/2\le K<n$.
\end{theorem}

We call the first alternative the YES case and the second the NO case, and the source promise excludes intermediate cover sizes. The range of $K$
excludes trivial cases: a cover in a cubic graph
must cover $3n/2$ edges, and each vertex covers at most three edges, so the size of the minimum vertex cover
is at least $n/2$. Moreover, a threshold $K\ge n$ is always a YES case.

Fix $\varepsilon$ as in Theorem~\ref{thm:cubic-source}. Throughout
the Finite-Finite reduction, the source input $H=(V_H,E_H)$ is cubic,
and its threshold $K$ satisfies
\begin{equation}
 n=|V_H|,\qquad m=|E_H|=3n/2,\qquad n/2\le K<n.
 \label{eq:source-range}
\end{equation}
Set $L:=6000$.
Create the following types, all with independent rate-one exponential
patience and unit match rewards $w_{ij} \equiv 1$ (Figure~\ref{fig:reduction}):
\begin{enumerate}
\item For each vertex $i\in V_H$, a supplier type $s_i$ of arrival rate $L$.
\item For each edge $e=\{i,j\}\in E_H$, a customer type $c_e$ of arrival
rate one, compatible with $s_i$ and $s_j$.
\item A \emph{flexible customer} type $c_0$, compatible with every supplier,
with arrival rate $\Gamma-m$, where
\begin{equation}
 \Gamma:=nL-(1+\varepsilon/2)K.
 \label{eq:Gamma}
\end{equation}
\end{enumerate}
The arrival rate of $c_0$ is well-defined as $\Gamma - m \geq \Gamma - 3n/2 \geq n(L-11/4) > 0$ since $\varepsilon \leq 1/2$, $K \leq n$, and $L=6000$. 
The compatibility graph is simple, connected, and bipartite, with
\begin{equation}
 |S|=n,\qquad |C|=3n/2+1,\qquad |E_G|=4n.
 \label{eq:image-size}
\end{equation}
Every supplier has degree four, and every edge customer has degree two.
The flexible type has degree $n$. Thus the total number of types and
compatibility edges is linear in the source size.

The intuition for constructing this instance is as follows:
$\Gamma$ is the total arrival rate of all customers and $nL$ is
the total arrival rate of all suppliers, so we have a supplier
surplus of $nL-\Gamma=(1+\varepsilon/2)K$. There are two cases:
\begin{enumerate}
\item On the one hand, we could take advantage of this surplus to match as many customers as possible. To do this, let $U^\star$ be a minimum vertex cover of $H$ and reserve up to
one waiting agent of each supplier type corresponding to a vertex
in $U^\star$ for edge customers. For instance, in
Figure~\ref{fig:reduction}, we reserve one agent of each type in
$\{s_1,s_2,s_4,s_6\}$. By doing this, we aim to keep compatible
suppliers available for edge customers while using additional
supply to serve flexible customers. 
For our parameter choices, this strategy matches nearly
all incoming customers when $|U^\star|=\tau(H)\le K$. In particular, the reserved suppliers abandon at a total
rate of at most $K$, so the supplier surplus of $(1+\varepsilon/2)K$ leaves an additional
margin of at least $\varepsilon K/2$. 

\item On the other hand, if $\tau(H)\ge(1+\varepsilon)K$, then this large minimum cover forces either many occupied supplier queues,
leading to supplier abandonment, or many edges with both supplier
endpoints empty, exposing arriving edge customers to abandonment
before a compatible supplier arrives. Combining these losses
gives an upper bound on throughput under every admissible policy strictly below the level
achievable in the first case. 
\end{enumerate}
Thus, under the source promise \eqref{eq:source-promise},
computing the optimal throughput to the accuracy specified below
would distinguish these two cases, which is an NP-hard task.

\paragraph{Fixed constants and source size.}

Throughout the Finite-Finite reduction, fix
\begin{equation}
\label{eq:constant-choice}
\varepsilon:=\frac1{100},
\qquad
L:=6000,
\qquad
n_0:=6\cdot10^{13},
\end{equation}
and define
\begin{equation}  \label{eq:sigma}
\sigma:=
\frac{\varepsilon}{8(L+1)}
-\frac{6(L+1)}{(L+4)(L-2)^2}.
\end{equation}
These constants are chosen for simplicity rather than an optimized
hardness gap; see Section~\ref{sec:questions}.
Direct substitution gives $\sigma>4\cdot10^{-8}>0$, and hence
\begin{equation}
\label{eq:cash-calibration}
\frac{2L}{\varepsilon}
\le \frac{\sigma n}{2}
\qquad(n\ge n_0).
\end{equation}
We assume $n\ge n_0$ without loss of generality: taking
$\lceil n_0/n\rceil$ disjoint copies of graph $H$ and scaling $K$
by the same factor preserves the source promise and increases
the instance size by at most a fixed factor.

\subsection{The fixed value gap and the NP-hardness reduction theorem}
\label{sec:value}
\label{ff:value-section}

The construction separates the optimal values of the two source cases
by a fixed fraction of the total arrival rate.

\begin{theorem}[Finite-Finite hardness]
\label{thm:ff-hardness}
There exists a fixed rational constant $\eta>0$ such that the
following promise problem is NP-hard: Given a Finite-Finite matching instance with unit rewards and rate-one
abandonment on both sides, together with a rational threshold
$b>0$, distinguish
\begin{equation}
 \frac{g^*}{\Lambda}\ge b+\eta
 \qquad\text{from}\qquad
 \frac{g^*}{\Lambda}\le b-\eta.
 \label{eq:value-promise}
\end{equation}
\end{theorem}
This theorem immediately implies the Finite-Finite part of
Theorem~\ref{thm:np_hardness}. In particular, choose
$0<\delta<\min\{1,2\eta\}$. Since $g^*\le\Lambda/2$, an estimate
$\widehat g$ satisfying $|\widehat g-g^*|\le\delta g^*$ also
satisfies
\[
 \left|\frac{\widehat g}{\Lambda}-\frac{g^*}{\Lambda}\right|
 \le\frac{\delta}{2}<\eta.
\]
Comparing $\widehat g/\Lambda$ with $b$ therefore solves the
promise problem. It remains to prove Theorem~\ref{thm:ff-hardness}, which is the focus of the rest of this section.

Following the intuition discussed earlier, we first upper bound the optimal throughput if the size of the minimum vertex cover is large.
\begin{proposition}[NO-instance bound] \label{prop:no-upper}
Suppose $\tau(H)\ge(1+\varepsilon)K$. Under the parameter choices in \eqref{eq:source-range},
\eqref{eq:sigma}, and \eqref{eq:cash-calibration}, every admissible policy satisfies
\begin{equation}  \label{eq:no-upper}
 r_\pi(T)\le\Gamma-\frac{\varepsilon K}{4(L+1)}
 \quad(T>0),
\end{equation}
hence
\[
g^*\le\Gamma-\frac{\varepsilon K}{4(L+1)}.
\]
\end{proposition}
The proof of the above proposition is presented in Section~\ref{ff:upper-section}. Now, we show the other direction: if the minimum vertex cover is small, then the optimal throughput is large. To do this, we explicitly construct a matching policy $\pi_U$ and analyze its performance to obtain a lower bound on the optimal throughput. The result is formalized below.
\begin{proposition}[YES-instance bound]  \label{comp:proposition}
Suppose $\tau(H)\le K$.
Under the parameter choices in \eqref{eq:source-range},
\eqref{eq:sigma}, and \eqref{eq:cash-calibration}, there exists a deterministic
stationary Markov policy $\pi_U$ such that
\begin{equation}
 g^{\pi_U}\ge
 \Gamma-n\left(\frac{\varepsilon}{8(L+1)}
                         -\frac{\sigma}{2}\right).
 \label{comp:gain}
\end{equation}
\end{proposition}
The proof of the above proposition is given in Section~\ref{ff:lower-section}, and here we prove Theorem~\ref{thm:ff-hardness} given the above two propositions.

\begin{proof}[Proof of Theorem~\ref{thm:ff-hardness}]
Define the two rational thresholds
\begin{equation}
 g_-:=\Gamma-\frac{\varepsilon K}{4(L+1)},\qquad
 g_+:=\Gamma-n\left(\frac{\varepsilon}{8(L+1)}
                              -\frac{\sigma}{2}\right).
 \label{eq:raw-thresholds}
\end{equation}
Set
\begin{equation}
 b_{H,K}:=\frac{g_++g_-}{2\Lambda},
 \qquad
 \eta:=\frac{\sigma}{8L}.
 \label{eq:value-promise-1}
\end{equation}
Since $K\ge n/2$ and $\Lambda<2nL$,
\begin{equation}
 \frac{g_+-g_-}{2\Lambda}
 =\frac{\varepsilon(K-n/2)}{8(L+1)\Lambda}
      +\frac{\sigma n}{4\Lambda}
 >\frac{\sigma}{8L}=\eta.
 \label{eq:raw-separation}
\end{equation}
If $\tau(H)\le K$, Proposition~\ref{comp:proposition} gives
\[
 \frac{g^*}{\Lambda}\ge\frac{g_+}{\Lambda}
 >b_{H,K}+\eta.
\]
If $\tau(H)\ge(1+\varepsilon)K$,
Proposition~\ref{prop:no-upper} gives
\[
 \frac{g^*}{\Lambda}\le\frac{g_-}{\Lambda}
 <b_{H,K}-\eta.
\]
Thus every produced instance satisfies the target promise.

Both the matching instance and the rational threshold $b_{H,K}$
have polynomial encoding length and can be computed in polynomial
time. An algorithm distinguishing the two target value ranges
would therefore distinguish the NP-hard source alternatives of
Theorem~\ref{thm:cubic-source}.
\end{proof}


The next two subsections establish the upper and lower bounds stated in Propositions~\ref{prop:no-upper} and \ref{comp:proposition}, respectively.

\subsection{A finite-horizon upper bound for every policy}  \label{ff:upper-section}
For $P\subseteq V_H$, let
\[
 u(P):=|\{e\in E_H:e\cap P=\varnothing\}|
\]
be the number of source edges uncovered by $P$. Define
$\Phi(P):=|P|+u(P)$. Note that adding one endpoint of every uncovered edge
extends $P$ to a vertex cover, which gives the following bound.
\begin{lemma}[Completing an occupied set]\label{lem:cover-completion}
For every $P\subseteq V_H$,
\begin{equation}
 \Phi(P)=|P|+u(P)\ge\tau(H).
 \label{eq:cover-completion}
\end{equation}
\end{lemma}
\begin{proof}
Start with $P$ and add one endpoint of every edge uncovered by $P$,
using a fixed ordering to choose that endpoint. Each originally covered
edge remains covered, and every other edge receives an endpoint. At most
$u(P)$ vertices are added, so the resulting cover has size at most
$|P|+u(P)$.
\end{proof} 

Recall that $Q_i^s(t)$ is the number of waiting suppliers at
vertex $i$, and define
\[
 P(t):=\{i\in V_H:Q_i^s(t)>0\}.
\]
The next two lemmas relate occupied supplier vertices to supplier
abandonment and uncovered edges to customer loss. Combining
these estimates with Lemma~\ref{lem:cover-completion} gives the
upper bound.

\subsubsection{Loss from occupied supplier vertices}


\begin{lemma}\label{lem:supplier-loss}
For every admissible policy $\pi$ and every $T>0$,
\begin{equation}
 \E M^\pi(T)\le nLT-\E \int_0^T|P(t)|\,dt.
 \label{eq:supplier-upper}
\end{equation}
\end{lemma}
\begin{proof}
Let $A_S(T)$ and $D_S(T)$ count supplier arrivals and abandonments, respectively, and note that
\[
 \sum_{i=1}^n Q^s_i(T) =A_S(T)-D_S(T)-M^\pi(T).
\]
Taking expectations on both sides, rearranging the terms, and dropping $\sum_{i=1}^n Q^s_i(T)$, we get
\[
\E M^\pi(T) \leq nLT - \E D_S(T) \overset{*}{=}  nLT - \E \int_0^T \sum_{i=1}^n Q^s_i(t) dt \leq nLT - \E \int_0^T |P(t)| dt,
\]
where $(*)$ follows because the expected number of abandonments equals the expected integral of the queue lengths, by the random time-change representation for Poisson processes, and the last inequality follows because the sum of queue lengths is bounded below by the number of nonempty queues.

\end{proof}
\subsubsection{Loss from uncovered edges}
\begin{lemma}\label{lem:customer-loss}
For every admissible policy $\pi$ and every $T>0$,
\begin{equation}
 \E M^\pi(T)\le\Gamma T-
 \frac{1}{1+2L}\E\int_0^T u(P(t))\,dt.
 \label{eq:customer-upper}
\end{equation}
\end{lemma}
\begin{proof}

Fix an admissible policy $\pi$, and let $\mathcal F_t$ denote the
observed history up to time $t$.
For each edge $e=\{i,j\}\in E_H$, let $N_e(t)$ be its rate-one
customer arrival process, with arrival times $\tau_{e,k}$, and define
\[
H_e(t):=\mathbf 1_{\{e\cap P(t-)=\varnothing\}}.
\]
Thus $H_e(\tau_{e,k})$ indicates that the arriving customer finds
both compatible supplier queues empty.

Let $Y_{e,k}$ indicate that this customer's patience expires before
the next supplier arrival at either endpoint of $e$.
Conditional on $\mathcal F_{\tau_{e,k}}$, the customer's patience
and the time to the next compatible supplier arrival are independent
exponential random variables with rates $1$ and $2L$, respectively.
Hence
\[
\mathbb E [Y_{e,k}\mid\mathcal F_{\tau_{e,k}}]
=\frac{1}{1+2L}.
\]

Let $B(T)$ count the customers arriving by $T$ who find both
endpoints empty and whose patience expires before the next
compatible supplier arrival. Then
\[
B(T)=\sum_{e\in E_H}\sum_{k\ge1}
\mathbf 1_{\{\tau_{e,k}\le T\}}
H_e(\tau_{e,k})Y_{e,k}.
\]
Since $\mathbf 1_{\{\tau_{e,k}\le T\}}H_e(\tau_{e,k})$ is
$\mathcal F_{\tau_{e,k}}$-measurable, the tower property and
Tonelli's theorem give
\begin{align*}
\mathbb E [B(T)]
&=
\sum_{e\in E_H}\sum_{k\ge1}
\mathbb E \!\left[
\mathbf 1_{\{\tau_{e,k}\le T\}}H_e(\tau_{e,k})
\mathbb E [Y_{e,k}\mid\mathcal F_{\tau_{e,k}}]
\right]\\
&=
\frac{1}{1+2L}\sum_{e\in E_H}
\mathbb E \!\left[\int_0^T H_e(t)\,dN_e(t)\right].
\end{align*}
The process $H_e$ is bounded and predictable, and $N_e$ has
intensity one. Applying the compensation formula therefore yields
\begin{align*}
\mathbb E [B(T)]
&=
\frac{1}{1+2L}\,\mathbb E \!\left[
\int_0^T\sum_{e\in E_H}H_e(t)\,dt
\right]=
\frac{1}{1+2L}\,
\mathbb E \!\left[\int_0^T u(P(t-))\,dt\right].
\end{align*}

No customer counted by $B(T)$ can ever be matched, regardless
of whether its expiration occurs before or after $T$.
Writing $A_C(T)$ for the total number of customer arrivals by $T$,
the empty initial state therefore implies
\[
M^\pi(T)\le A_C(T)-B(T).
\]
Since $\mathbb E [A_C(T)]=\Gamma T$, we conclude that
\[
\mathbb E [M^\pi(T)]
\le
\Gamma T-\frac{1}{1+2L}\,
\mathbb E \!\left[\int_0^T u(P(t))\,dt\right].
\]
Here $P(t-)$ may be replaced by $P(t)$ in the time integral because
the process has only finitely many events on every bounded interval.
\end{proof}

\subsubsection{Balancing the losses}  \label{ff:balancing-section}

Supplier abandonment bounds how many vertices a policy can keep
occupied, and customer loss bounds how many edges it can leave uncovered. Thus,
adding these two bounds controls the size of a cover obtained from the
occupied vertices.

\begin{lemma}[Finite-horizon upper bound]\label{lem:soundness}
For every admissible policy $\pi$ and every $T>0$,
\begin{equation}
 \frac1T\E \int_0^T\Phi(P(t))\,dt
 \le (1+\varepsilon/2)K+2(L+1)\bigl(\Gamma-r_\pi(T)\bigr).
 \label{eq:cover-decoding-bound}
\end{equation}
Consequently,
\begin{equation}
 r_\pi(T)\le\Gamma-
 \frac{\tau(H)-(1+\varepsilon/2)K}{2(L+1)}.
 \label{eq:soundness}
\end{equation}
\end{lemma}
\begin{proof}
Rearrange the supplier and customer conservation inequalities and add them to get
\begin{align*}
 \frac1T\E \int_0^T\bigl(|P(t)|+u(P(t))\bigr)\,dt
 &\le nL-r_\pi(T)+(1+2L)\bigl(\Gamma-r_\pi(T)\bigr)\\
 &=(1+\varepsilon/2)K+2(L+1)\bigl(\Gamma-r_\pi(T)\bigr),
\end{align*}
where $nL-\Gamma=(1+\varepsilon/2)K$.
This proves \eqref{eq:cover-decoding-bound}.
Lemma~\ref{lem:cover-completion} bounds the integrand below by
$\tau(H)$, giving \eqref{eq:soundness}.
\end{proof}

\begin{proof}[Proof of Proposition~\ref{prop:no-upper}]
Substitute $\tau(H)\ge(1+\varepsilon)K$ into~\eqref{eq:soundness}
and then take the liminf and supremum in the definition of $g^*$.
\end{proof}

\subsection{A small cover serves nearly all demand}  \label{ff:lower-section}

The upper bound associates a cover with the supplier vertices occupied by a
policy.  We now establish the converse direction of the reduction: a
small vertex cover $U$ specifies where a stationary Markov policy $\pi_U$ should keep inventory.

\subsubsection{The cover policy $\pi_U$}  \label{comp:policy}

We now specify the cover policy $\pi_U$.
Since $\tau(H)\le K$, enlarge a vertex cover, if necessary, to
obtain $U\subseteq V_H$ with $|U|=K$.

For every edge $e\in E_H$, fix one endpoint $o(e)\in e\cap U$.
Under $\pi_U$, a customer of type $c_e$ is matched only
to a supplier of type $s_{o(e)}$, as soon as both are available.
We say that this customer of type $c_e$ is assigned to vertex
$o(e)$.
These matches take
priority over flexible-customer matches: flexible customers are
matched as soon as a supplier is available outside $U$ or a
supplier queue in $U$ contains at least two agents. Unmatched
agents remain in their queues until matched or until their
patience expires. All ties are resolved using fixed queue
orderings, with individuals within each queue served in arrival
order.

For each vertex $i\in V_H$, define
\[
 \mathcal F_i:=\{e\in E_H:o(e)=i\},\qquad
 k_i:=|\mathcal F_i|.
\]
The edge customers assigned to vertex $i$ are those of
types $c_e$ with $e\in\mathcal F_i$. Since each such type has
arrival rate one, their total arrival rate is $k_i$.
In particular, $k_i=0$ outside $U$, $k_i\le3$ for every $i$ since the source graph is cubic, and
$\sum_{i\in V_H} k_i=m$ since $U$ is a vertex cover.

Now that the cover policy is completely defined, we present two key lemmas that contribute to proving Proposition~\ref{comp:proposition}. First, recall that $Q_e^c$ is the number of waiting customers of type $c_e$, $Q_0^c$
is the number of waiting flexible customers, and $Q_i^s$ is the number of waiting suppliers. 
By Lemma~\ref{comp:recurrence}, the full queue-length
process under $\pi_U$ has a unique stationary distribution
on the class reachable from empty. Denote this distribution
by $\nu_U$, and write $\E_{\nu_U}$ for expectation under it.

We next present a lemma that bounds the backlog of edge customers. Its proof is deferred to Section~\ref{comp:edge-section}.
\begin{lemma}[Edge-customer backlog]  \label{comp:edge-backlog}
It holds that
\begin{equation}
 \sum_{e\in E_H}\E_{\nu_U}Q_e^c
 \le n\left(\frac{\varepsilon}{8(L+1)}-\sigma\right).
 \label{comp:edge-total}
\end{equation}
\end{lemma}
The following provides intuition for the above bound. 
At a selected vertex, the supplier arrival rate $L=6000$ is large compared
with the assigned customer arrival rate $k_i\le3$.  Moreover, the reservation
prevents flexible customers from removing the last supplier.  As a
result, an assigned customer waits only during a short shortage of
supply, which allows us to bound the expected number of waiting edge customers in the system. Next, the following lemma bounds the backlog of flexible customers.
Its proof is deferred to
Section~\ref{comp:flexible-section}.
\begin{lemma}[Flexible-customer backlog]
\label{comp:lem-flexible-backlog}
It holds that
\begin{equation}  \label{comp:eq-flexible-backlog}
\mathbb E_{\nu_U}Q_0^c
\le \frac{\sigma n}{2}.
\end{equation}
\end{lemma}
To understand this bound, observe that whenever flexible customers are waiting, at most $K$ suppliers
are present, so the supplier abandonment rate is at most $K$.
The supplier arrival rate therefore exceeds the total customer
arrival rate plus supplier abandonment by at least
$nL-\Gamma-K=\varepsilon K/2$. We then present a Lyapunov argument that uses this restoring
drift to bound the expected flexible-customer queue length.


We are now ready to prove Proposition~\ref{comp:proposition} below.
\begin{proof}[Proof of Proposition~\ref{comp:proposition}]

Let $Q^{\rm tot}(t):=\sum_eQ_e^c(t)+Q_0^c(t)$ be the total number of waiting
customers at time $t$. Let $\mathcal A$ be the generator of the full queue-length process
under $\pi_U$, and let $r_U(q)$ denote its instantaneous matching
rate in state $q$. For
$f(q):=\sum_{e\in E_H}q_e^c+q_0^c$, customer arrivals, abandonment,
and matches give
\[
(\mathcal A f)(q)=\Gamma-f(q)-r_U(q).
\]
The stationary moment bounds in Lemma~\ref{comp:recurrence}
justify applying the stationary generator identity to $f$
(by truncation), yielding
\[
0=\E_{\nu_U}[\mathcal A f]
 =\Gamma-\E_{\nu_U}[Q^{\rm tot}]
         -\E_{\nu_U}[r_U(Q)].
\]
Since $r_U$ is bounded by the total arrival rate $\Lambda$,
convergence from the empty state to $\nu_U$ and Cesàro averaging
give $g^{\pi_U}=\E_{\nu_U}[r_U(Q)]$. Consequently,
\[
g^{\pi_U}
=\Gamma-\E_{\nu_U}[Q^{\rm tot}]
=\Gamma-\sum_{e\in E_H}\E_{\nu_U}[Q_e^c]
       -\E_{\nu_U}[Q_0^c].
\]

Finally, the bounds in \eqref{comp:edge-total} and
\eqref{comp:eq-flexible-backlog} give
\[
\sum_{e\in E_H}\E_{\nu_U}[Q_e^c]
\le
n\left(\frac{\varepsilon}{8(L+1)}-\sigma\right),
\qquad
\E_{\nu_U}[Q_0^c]\le\frac{\sigma n}{2}.
\]
Substituting these bounds yields
\[
g^{\pi_U}
\ge
\Gamma
-n\left(\frac{\varepsilon}{8(L+1)}-\sigma\right)
-\frac{\sigma n}{2}
=
\Gamma
-n\left(\frac{\varepsilon}{8(L+1)}-\frac{\sigma}{2}\right),
\]
completing the proof.
\end{proof}

It remains to prove Lemmas~\ref{comp:edge-backlog}
and~\ref{comp:lem-flexible-backlog}. Define
\[
 Q_i^{c,\text{tot}}(t)
 :=\sum_{e\in\mathcal F_i}Q_e^c(t),
 \qquad
 h_U(Q^s(t))
 :=\sum_{i\notin U}Q_i^s(t)
       +\sum_{i\in U}(Q_i^s(t)-1)^+.
\]
Here $Q_i^{c,\text{tot}}$ is the total number of waiting edge
customers served at vertex $i$, and $h_U(Q^s)$ counts suppliers
eligible to serve flexible customers. Starting empty, the policy
maintains
\begin{equation}
 Q_i^sQ_i^{c,\text{tot}}=0
 \quad(i\in V_H),\qquad
 Q_0^ch_U(Q^s)=0.
 \label{comp:complementarity}
\end{equation}
The first relation holds because waiting edge customers are
matched whenever their designated supplier queue becomes
nonempty. The second holds because waiting flexible customers
are matched whenever eligible inventory becomes positive.

\subsubsection{The assigned edge-customer backlog}  \label{comp:edge-section}

\begin{proof}[Proof of Lemma~\ref{comp:edge-backlog}]
Let $\mathcal A$ denote the generator of the full queue-length
process under $\pi_U$. Fix $i\in U$ and set $Z_i:=Q_i^s-Q_i^{c,\text{tot}}$. By \eqref{comp:complementarity},
\[
Q_i^s=(Z_i)^+,
\qquad
Q_i^{c,\text{tot}}=(-Z_i)^+.
\]
Now, consider the nonnegative
test function
\[
V_i(q)
:=
\frac12 q_i^{c,\text{tot}}
       \bigl(q_i^{c,\text{tot}}+1\bigr)
+\frac{k_i}{L+1}(1-Z_i(q))^+.
\]
The first term is the usual quadratic Lyapunov function used to bound the expectation of $Q_i^{c, \text{tot}}$. The second term is introduced to distinguish the states with $Z_i = 0$ and $Z_i = 1$. In both cases, $Q_i^{c, \text{tot}} = 0$, but the latter state has a supplier waiting in the system. We compute the drift by considering how many suppliers are waiting
at vertex $i$.

\emph{Case 1: No suppliers are waiting ($Z_i(q)=-r\le0$).}
By \eqref{comp:complementarity}, exactly $r$ customers whose types are indexed by $\mathcal F_i$
are waiting, i.e., $Q_i^{c, \text{tot}} = r$.  So the drift is given by
\[
(\mathcal A V_i)(q)
=
k_i\left(r+1+\frac{k_i}{L+1}\right)
-(L+r)\left(r+\frac{k_i}{L+1}\right).
\]
The first term corresponds to an edge-customer arrival at rate
$k_i$, which increases $Q_i^{c,\text{tot}}$ by one.
The second term corresponds to supplier arrivals and customer
abandonments. If $r\ge1$, these events occur at total rate $L+r$
and reduce $Q_i^{c,\text{tot}}$ by one, changing $V_i$ by
$-(r+k_i/(L+1))$. If $r=0$, only a supplier arrival is possible;
it changes $Z_i$ from zero to one and decreases $V_i$ by
$k_i/(L+1)$. Thus the same drift formula holds at $r=0$.

\emph{Case 2: Exactly one supplier is waiting ($Z_i(q)=1$).}
We have $Q_i^{c, \text{tot}} = 0$, so $V_i(q)=0$. The drift in this case is
\[
(\mathcal A V_i)(q)=\frac{k_i(k_i+1)}{L+1}.
\]
The above term corresponds to either the abandonment of a supplier of type $i$ or the arrival of a customer whose type is indexed by $\mathcal F_i$, each of which increases $V_i$ by $k_i/(L+1)$. The rate of this transition is $1 + k_i$.

\emph{Case 3: At least two suppliers are waiting ($Z_i(q)\ge2$).}
The function $V_i$ is zero whenever $Z_i\ge1$.
Since a single transition changes $Z_i$ by at most one,
it remains in this region after the transition. Hence $
(\mathcal A V_i)(q)=0$.

Combining the three cases and using
$Q_i^{c,\text{tot}}=(-Z_i(q))^+$ gives
\[
(\mathcal A V_i)(q)
=
\frac{k_i(k_i+1)}{L+1}\ind{Z_i(q)\le1}
-\left(L-k_i+\frac{k_i}{L+1}\right)q_i^{c,\text{tot}}
-\bigl(q_i^{c,\text{tot}}\bigr)^2.
\]
Since $q_i^{c,\text{tot}}$ is a nonnegative integer,
$\bigl(q_i^{c,\text{tot}}\bigr)^2\ge q_i^{c,\text{tot}}$.
Therefore,
\[
(\mathcal A V_i)(q)
\le
\frac{k_i(k_i+1)}{L+1}
-\left(L+1-k_i+\frac{k_i}{L+1}\right)q_i^{c,\text{tot}}.
\]

The stationary second-moment bounds in
Lemma~\ref{comp:recurrence} justify
$\E_{\nu_U}[\mathcal A V_i]=0$.
Taking stationary expectations and rearranging yields
\begin{equation}
\label{comp:local-backlog}
\E_{\nu_U}[Q_i^{c,\text{tot}}]
\le
\frac{k_i(k_i+1)}{(L+1)^2-k_iL}.
\end{equation}
Since $0\le k_i\le3$, this implies
\[
\E_{\nu_U}[Q_i^{c,\text{tot}}]
\le
\frac{4k_i}{L^2-L+1}.
\]
Note that one could also use a linear function of $Q_i^{c, \text{tot}}$ as the test function instead of a quadratic function, but that gives a weaker bound of $\E_{\nu_U}[Q_i^{c,\text{tot}}] = O(L^{-1})$, whereas a carefully constructed quadratic test function gives the tighter bound of $\E_{\nu_U}[Q_i^{c,\text{tot}}] = O(L^{-2})$.
Intuitively, supply shortages are rare when $L$ is large,
and supplier arrivals replenish inventory on the $1/L$
time scale. The quadratic drift calculation above captures
both effects and yields the $O(L^{-2})$ backlog bound.

Every edge-customer type is assigned to exactly one vertex in $U$,
and $\sum_{i\in U}k_i=m=3n/2$. Summing over $i\in U$ gives
\begin{align*}
\sum_{e\in E_H}\E_{\nu_U}[Q_e^c]
&=
\sum_{i\in U}\E_{\nu_U}[Q_i^{c,\text{tot}}]\\
&\le
\frac{4m}{L^2-L+1}
=
\frac{6n}{L^2-L+1}\\
&\le
n\,\frac{6(L+1)}{(L+4)(L-2)^2}\\
&=
n\left(\frac{\varepsilon}{8(L+1)}-\sigma\right).
\end{align*}
The last inequality follows from $L>2$ and
\[
(L+1)(L^2-L+1)-(L+4)(L-2)^2=12L-15>0,
\]
and the final equality uses the definition of $\sigma$
in \eqref{eq:sigma}.
\end{proof}

\subsubsection{The flexible-customer backlog}  \label{comp:flexible-section}
\begin{proof}[Proof of Lemma~\ref{comp:lem-flexible-backlog}]
Define the supplier--customer imbalance and total population by
\[
Z(t):=\sum_{i\in V_H}Q_i^s(t)
      -\sum_{e\in E_H}Q_e^c(t)-Q_0^c(t),
\qquad
N(t):=\sum_{i\in V_H}Q_i^s(t)
      +\sum_{e\in E_H}Q_e^c(t)+Q_0^c(t).
\]
For a queue state $q$, write $Z(q)$ and $N(q)$ for the
corresponding functions of that state. Let $\mathcal A$ denote the generator of the full queue-length
process under $\pi_U$, and consider the test function
\[
V(q):=\bigl(Z(q)-(nL-\Gamma)\bigr)^2+N(q).
\]
Matches leave $Z$ unchanged. Supplier arrivals and customer
abandonment increase $Z$ by one, while customer arrivals
and supplier abandonment decrease it by one. Since every
waiting agent abandons at rate one,
\[
(\mathcal A Z)(q)=nL-\Gamma-Z(q),
\]
and the total rate of these unit jumps is $\Lambda+N(q)$.
Consequently,
\[
\mathcal A\!\left[(Z-(nL-\Gamma))^2\right](q)
=
-2\bigl(Z(q)-(nL-\Gamma)\bigr)^2+\Lambda+N(q).
\]
Also, arrivals add at most one agent, each abandonment removes
one agent, and matches can only reduce the total population.
Therefore,
\[
(\mathcal A N)(q)\le\Lambda-N(q).
\]
Adding these two drift bounds cancels the population term and gives
\begin{align*}
(\mathcal A V)(q)
&\le
2\Lambda-2\bigl(Z(q)-(nL-\Gamma)\bigr)^2\\
&\le
2\Lambda-4\varepsilon K\,q_0^c,
\end{align*}
where the last inequality follows from the following observation:
if $q_0^c>0$, \eqref{comp:complementarity} implies
$h_U(q^s)=0$: every supplier queue outside $U$ is empty,
and each queue in $U$ contains at most one supplier.
Thus the total supplier inventory is at most $K$, giving
\[
Z(q)\le K-q_0^c.
\]
Since $nL-\Gamma=(1+\varepsilon/2)K$, it follows that
\[
\bigl(Z(q)-(nL-\Gamma)\bigr)^2
\ge
\left(\frac{\varepsilon K}{2}+q_0^c\right)^2
\ge
2\varepsilon K\,q_0^c.
\]
Now, the stationary second-moment bounds in
Lemma~\ref{comp:recurrence} justify applying the stationary
generator identity to $V$ by truncation. Taking expectations
under $\nu_U$ therefore yields
\[
0=\E_{\nu_U}[\mathcal A V]
\le
2\Lambda-4\varepsilon K\,\E_{\nu_U}[Q_0^c].
\]
Rearranging, we obtain
\[
\E_{\nu_U}[Q_0^c]
\le
\frac{\Lambda}{2\varepsilon K}
\le
\frac{2L}{\varepsilon}
\le
\frac{\sigma n}{2}.
\]
Here the second inequality uses $\Lambda<2nL$ and $K\ge n/2$,
and the final inequality follows from
$n\ge n_0\ge4L/(\varepsilon\sigma)$.
This proves \eqref{comp:eq-flexible-backlog}.
\end{proof}

\section{Consequences for policy design}  \label{sec:policy-consequences}

In this section we prove Corollaries~\ref{cor:policy-hardness}
and~\ref{cor:deployment}, which establish the hardness of constructing
near-optimal policies under the stated execution and convergence
requirements. Both proofs use the reduction underlying
Theorem~\ref{thm:np_hardness}.



Let $g_-<g_+$ be the thresholds in \eqref{eq:raw-thresholds}, and write their gap as
$\Delta:=g_+-g_-$. Recall that
\begin{equation}  \label{dep:parameters}
 g_- =\Gamma-\frac{\varepsilon K}{4(L+1)},\qquad
 g_+ =\Gamma-n\left(\frac{\varepsilon}{8(L+1)}
                              -\frac{\sigma}{2}\right), \qquad
 \Delta\ge\frac{\sigma n}{2}>0.
\end{equation}

\subsection{Recovering a vertex cover from the queues}

For every edge $e\in E_H$, choose an endpoint $v_e\in e$
using a fixed vertex ordering. For $P\subseteq V_H$, define
\begin{equation}
 \mathcal C(P)
 :=P\cup\{v_e:e\in E_H,\ e\cap P=\varnothing\}.
 \label{eq:C(P)-def}
\end{equation}
By definition, this is a vertex cover. It is clear that its construction takes polynomial time. 

\begin{lemma}[A queue observation reveals a cover] \label{dep:decoding}
Let $U_T$ be uniform on $[0,T]$, independently of the matching process.
Let $P(t)=\{i\in V_H:Q_i^s(t)>0\}$ be the occupied supplier vertices at time $t$, and $r_\pi(T)=\E [M^\pi(T)]/T$ be the expected number of matches per unit time under policy $\pi$ up to time $T$.
Then for every admissible policy $\pi$ and every $T>0$, 
\begin{equation}
 \E |\mathcal C(P(U_T))|
 \le (1+\varepsilon/2)K+2(L+1)(\Gamma-r_\pi(T)).
 \label{dep:decoding-mean}
\end{equation}
Thus if $r_\pi(T)\ge(g_-+g_+)/2$, then
\begin{equation}
 \Prob\{|\mathcal C(P(U_T))|<(1+\varepsilon)K\}
 \ge p_*:=\frac{(L+1)\sigma}{2(1+\varepsilon)}>0.
 \label{dep:decoding-probability}
\end{equation}
\end{lemma}

Here is the implication of this lemma: since on a NO source instance every cover has size at least
$(1+\varepsilon)K$, a sampled $|\mathcal C(P(U_T))|$ below $(1+\varepsilon)K$ under the promise immediately tells us that $\tau(H) \le K$, hence certifies a YES answer. In other words, a policy with sufficiently good finite-horizon performance has a fixed positive probability of correctly identifying a YES instance and never incorrectly answers YES to a NO instance. This will play a key role in our two subsequent corollaries involving the class RP.

\begin{proof}
For $P\subseteq V_H$, let $u(P)=|\{e\in E_H:e\cap P=\varnothing\}|$.
By the definition of $U_T$, we have
\[
 \E[|P(U_T)|+u(P(U_T))]
 =\frac1T\E\int_0^T (|P(t)|+u(P(t)))\,dt.
\]
Recall \eqref{eq:cover-decoding-bound} from Lemma~\ref{lem:soundness}: 
\[
\frac1T\E \int_0^T(|P(t)|+u(P(t))) \,dt
 \le (1+\varepsilon/2)K+2(L+1)\bigl(\Gamma-r_\pi(T)\bigr)
\]
Combining the bound $|\mathcal C(P(t))| \le |P(t)|+u(P(t))$ with
\eqref{eq:cover-decoding-bound} proves
\eqref{dep:decoding-mean}.

If $r_\pi(T)\ge(g_-+g_+)/2$, then $\Delta=g_+-g_-$ and
the definition of $g_-$ in \eqref{dep:parameters} give
\[
 \Gamma-r_\pi(T)
 \le \Gamma-g_- -\frac{\Delta}{2}
 =\frac{\varepsilon K}{4(L+1)}-\frac{\Delta}{2}.
\]
Substituting this into \eqref{dep:decoding-mean} then yields
\begin{align*}
 \E |\mathcal C(P(U_T))|
 &\le (1+\varepsilon/2)K
       +2(L+1)\bigl(\Gamma-r_\pi(T)\bigr)\\
 &\le (1+\varepsilon/2)K
       +2(L+1)\left(
          \frac{\varepsilon K}{4(L+1)}-\frac{\Delta}{2}
       \right)\\
 &= (1+\varepsilon)K-(L+1)\Delta.
\end{align*}
Markov's inequality therefore gives
\[
 \Prob\{|\mathcal C(P(U_T))|<(1+\varepsilon)K\}
 \ge \frac{(L+1)\Delta}{(1+\varepsilon)K}
 \ge \frac{(L+1)\sigma}{2(1+\varepsilon)},
\]
where the last inequality uses $K<n$ and $\Delta\ge\sigma n/2$.
\end{proof}

\subsection{Proofs of Corollaries~\ref{cor:policy-hardness} and \ref{cor:deployment}}  \label{sec:cor-proofs}

Both corollaries hold with
\[
 \eta_0:=\frac{\sigma}{16L},
 \qquad
 \rho:=\frac{p_*}{4},
\]
where $\sigma$ is defined in \eqref{eq:sigma} and $p_*$
in Lemma~\ref{dep:decoding}.

Lemma~\ref{dep:decoding} recovers a vertex cover $\mathcal C(P(U_T))$ from the occupied
supplier set at a uniformly chosen time $U_T$. The following lemma
allows us to compute in polynomial time an approximation to that observation.
Its proof is deferred to Appendix~\ref{app:technical-details}.

\begin{lemma}[Sampling the occupied supplier set]
\label{dep:execution}
Fix a rational $0<\rho<1$. Given a Finite-Finite reduction
instance, an admissible efficiently executable policy $\pi$ as described in Section~\ref{subsec:NP-hardness},
and an integer $T\ge1$, there is a randomized algorithm that returns
$\widehat P\subseteq V_H$ satisfying
\[
 \bigl\|\mathcal L(\widehat P)
       -\mathcal L(P(U_T))\bigr\|_{\mathrm{TV}}
 \le\rho,
\]
in worst-case time polynomial in the instance length, the
policy-description length, and the numerical value of $T$. Here $U_T$ is uniform
on $[0,T]$ and independent of the matching process.
\end{lemma}

We are now ready to prove Corollaries~\ref{cor:policy-hardness} and \ref{cor:deployment} below.

\begin{proof}[Proof of Corollary~\ref{cor:policy-hardness}]
Consider a YES source instance and suppose a policy $\pi$
satisfies both guarantees in the claim of the corollary: $g^\pi \ge g^*-\eta_0\Lambda$ and $t_{\mathrm{mix}}^{0,\pi}(\rho) \le p(B)$. 
Since for a YES instance $g^*\ge g_+$ by Proposition \ref{comp:proposition}, and
\[
\eta_0\Lambda < \frac{\sigma}{16L} (2nL) \le \frac{\Delta}{4},
\]
we deduce that
\[
 g^\pi\ge g_+-\eta_0\Lambda> g_+ - \frac{\Delta}{4} > \frac{g_-+g_+}{2},
\]
where the last inequality uses $\Delta=g_+-g_-$ and $g_+>g_-$.

Thus, by the definition of $g^\pi$ as a liminf,
$r_\pi(T)>(g_-+g_+)/2$ for all sufficiently large $T$.
Lemma~\ref{dep:decoding} therefore gives
\[
 \Prob \{|\mathcal C(P(U_T))|<(1+\varepsilon)K\}\ge p_*
\]
for all such $T$, where $U_T$ is independent and uniform on $[0,T]$.
Write
\[
 \Prob \{|\mathcal C(P(U_T))|<(1+\varepsilon)K\}
 =
 \frac1T\int_0^T
 \Prob \{|\mathcal C(P(s))|<(1+\varepsilon)K\}\,ds.
\]
Since the queue law converges to $\nu_\pi$ in total variation by Lemma~\ref{comp:recurrence},
the integrand converges to the corresponding stationary probability.
Its time average has the same limit, so
\begin{equation}
 \Prob_{\nu_\pi}\{|\mathcal C(P)|<(1+\varepsilon)K\}\ge p_*.
 \label{mix:stationary-certificate}
\end{equation}

We now use the mixing guarantee to obtain such a cover within
a polynomial horizon. Set
\[
 T:=\left\lceil\frac{p(B)+1}{\rho}\right\rceil.
\]
For every $s\ge p(B)+1$, the definition of the mixing time ensures that the queue law is within total-variation
distance $\rho$ of $\nu_\pi$. Together with
\eqref{mix:stationary-certificate}, this implies
\[
 \Prob \{|\mathcal C(P(s))|<(1+\varepsilon)K\}
 \ge p_*-\rho.
\]
Moreover, $(p(B)+1)/T\le\rho$, so these times occupy at least
a $1-\rho$ fraction of $[0,T]$. Averaging over $U_T$ gives
\[
 \Prob \{|\mathcal C(P(U_T))|<(1+\varepsilon)K\}
 \ge (1-\rho)(p_*-\rho)
 \ge p_*-2\rho.
\]

We now apply Lemma~\ref{dep:execution} at accuracy $\rho$ to obtain
$\widehat P$, and complete it to the cover $\mathcal C(\widehat P)$.
The probabilities of the decoded-cover event under $\widehat P$
and $P(U_T)$ differ by at most $\rho$. Recalling that
$\rho=p_*/4$, we obtain
\begin{align*}
 \Prob\{|\mathcal C(\widehat P)|<(1+\varepsilon)K\}
 &\ge
 \Prob \{|\mathcal C(P(U_T))|<(1+\varepsilon)K\}-\rho\\
 &\ge p_*-3\rho=\frac{p_*}{4}.
\end{align*}
This bound holds conditional on any returned policy satisfying
both guarantees in the statement of the corollary. Since policy construction succeeds with
probability at least $2/3$, one trial produces such a cover
with probability at least $(2/3)(p_*/4)=p_*/6$ on a YES instance.

Now consider the following strategy: accept exactly when the decoded cover has fewer than
$(1+\varepsilon)K$ vertices. A NO instance is never accepted:
the completion always produces a genuine vertex cover, and
every vertex cover of a NO instance has size at least
$(1+\varepsilon)K$. A constant number of independent trials,
accepting if any trial accepts, raises the YES acceptance
probability to at least $2/3$.

Finally, policy construction takes polynomial time and produces a
polynomial-length description. Since $\rho$ is fixed and
$T$ is polynomial in $B$, Lemma~\ref{dep:execution} also gives
polynomial worst-case execution time; cover completion is
polynomial as well. We have therefore obtained a one-sided
randomized polynomial-time solver for the NP-hard source
promise, implying $\mathrm{NP}=\mathrm{RP}$.
\end{proof}

\begin{proof}[Proof of Corollary~\ref{cor:deployment}]
Given a source instance, form its matching instance and use the
assumed algorithm to construct a policy $\pi$. Set
\[
 T:=\max\{1,\lceil p(B)\rceil\}.
\]
Consider a YES source instance and condition on a returned policy
satisfying both guarantees in the corollary. Since $T\ge p(B)$
and $g^*\ge g_+$ by Proposition~\ref{comp:proposition},
\begin{align*}
 r_\pi(T)
 &\ge g^\pi-\eta_0\Lambda
 \ge g^*-2\eta_0\Lambda\\
 &\ge g_+-2\eta_0\Lambda
 >g_+-\frac{\sigma n}{4}\\
 &\ge g_+-\frac{\Delta}{2}
 =\frac{g_-+g_+}{2}.
\end{align*}
Here the last two inequalities use
$\eta_0\Lambda<\sigma n/8$ and $\Delta\ge\sigma n/2$,
as given by \eqref{dep:parameters}.
Lemma~\ref{dep:decoding} therefore yields
\[
 \Prob \{|\mathcal C(P(U_T))|<(1+\varepsilon)K\}\ge p_*,
\]
where $U_T$ is independent and uniform on $[0,T]$.

Apply Lemma~\ref{dep:execution} at accuracy $\rho:=p_*/4$
to obtain $\widehat P$, and complete it to the cover
$\mathcal C(\widehat P)$. The probabilities of the decoded-cover
event under $\widehat P$ and $P(U_T)$ differ by at most $\rho$.
Consequently,
\begin{align*}
 \Prob\{|\mathcal C(\widehat P)|<(1+\varepsilon)K\}
 &\ge
 \Prob \{|\mathcal C(P(U_T))|<(1+\varepsilon)K\}-\rho\\
 &\ge p_*-\rho=\frac{3p_*}{4}.
\end{align*}
This bound holds conditional on any returned policy satisfying
both guarantees. Since policy construction succeeds with
probability at least $2/3$, one trial produces such a cover
with probability at least $(2/3)(3p_*/4)=p_*/2$ on a YES instance.

Now consider the following strategy: accept exactly when the decoded cover has fewer than
$(1+\varepsilon)K$ vertices. A NO instance is never accepted:
completion always produces a genuine vertex cover, and every
vertex cover of a NO instance has size at least
$(1+\varepsilon)K$. A constant number of independent trials,
accepting if any trial finds such a cover, raises the YES
acceptance probability to at least $2/3$.

Policy construction takes polynomial time and produces a
polynomial-length description. Since $\rho$ is fixed and
$T$ is polynomial in $B$, Lemma~\ref{dep:execution} gives
polynomial worst-case execution time; cover completion is
polynomial as well. This produces a one-sided randomized
polynomial-time solver for the NP-hard source promise,
implying $\mathrm{NP}=\mathrm{RP}$.
\end{proof}

\section{Open questions and future directions}  \label{sec:questions}

The classification identifies the tractable infinite-patience
regimes and establishes value hardness in the Finite-Finite
and Finite-Zero regimes. It leaves two principal computational
questions.

\paragraph{The sharp polynomial-time approximation ratio.}
How closely can a polynomial-time algorithm approximate optimal online
reward in each hard regime? On the one hand,
we have obtained a fixed relative-error
obstruction for numerical value approximation. With the current
constants in \eqref{eq:constant-choice}, approximating the optimal
value within relative error $10^{-12}$ is NP-hard, so the
excluded approximation factor is extremely close to one.
On the other hand, the policy guarantees in
Table~\ref{tab:online-ratios} give $1/2$ for Finite-Finite matching
and $1-1/e+\delta_0$ for Finite-Zero matching, even with weighted
rewards. The unit-reward, common-abandonment subclass may admit
better algorithms. Determining a sharp factor therefore requires progress
in two directions: (i) stronger hardness bounds, and
(ii) improved approximation algorithms.

For (i), the present reduction loses much of the source approximation
gap in a large common reward term. Its relative separation
is 
\[
 \frac{g_+-g_-}{g_+}=O(\varepsilon/L^2),
\]
whereas the source vertex-cover sizes differ by a factor
$1+\varepsilon$. We need to choose $L$ large so that the constant $\sigma$ in \eqref{eq:sigma} is positive. The inverse-square dependence from the source gap reflects a limitation of the current construction and bounds. A reduction that preserves a larger fraction of the
source gap could therefore yield a stronger hardness
factor.

\paragraph{The exact complexity of the hard value promises.}
Our reductions establish NP-hardness but do not establish membership
in NP for either the Finite-Finite or the Finite-Zero value promise.
Can YES instances at a fixed normalized gap admit
polynomial-length certificates that can be checked in polynomial
time? Alternatively, does the problem admit a polynomial-space
algorithm, and can one prove PSPACE-hardness? NP-completeness and
PSPACE-completeness are natural possibilities to investigate,
although our results do not rule out other complexity classifications.
Determining the exact complexity of matching with abandonment would
clarify its computational limits and guide algorithm design.

\clearpage

\appendix

\section{Proof of Theorem~\ref{thm:permanent} on infinite patience}   \label{app:positive}

Compute an optimal flow $f$ in \eqref{eq:flow-lp}.
For each edge $e=(i,j)$ with $f_e:=f_{ij}>0$, choose one
infinitely patient endpoint for storage and maintain an
initially empty inventory $Q_e$ there.
The policy $\pi$ operates as follows.
\begin{enumerate}
\item Each supplier arrival of type $i$ chooses edge $(i,j)$
with probability $f_{ij}/\lambda_i$; each customer arrival of
type $j$ chooses edge $(i,j)$ with probability
$f_{ij}/\gamma_j$. No edge is chosen with the remaining
probability. All choices are independent across arrivals
and of the arrival and patience processes.
The flow constraints make these probabilities valid.

\item If edge $e$ is chosen, an arrival at its storage endpoint
joins $Q_e$. An arrival at its other endpoint matches
immediately with one agent in $Q_e$ if $Q_e>0$.
\end{enumerate}

Agents for whom no edge is chosen, and arrivals at the other
endpoint that find an empty inventory, are never subsequently
matched. They remain admitted until their natural departure;
infinitely patient agents remain indefinitely.

\begin{proof}[Proof of Theorem~\ref{thm:permanent}]
For any admissible policy $\pi'$, each agent participates in
at most one match. Starting empty therefore gives
\[
 \sum_j\E M_{ij}^{\pi'}(T)\le\lambda_iT,
 \qquad
 \sum_i\E M_{ij}^{\pi'}(T)\le\gamma_jT.
\]
Thus
$(\E M_{ij}^{\pi'}(T)/T)_{(i,j)\in E_G}$
is feasible in \eqref{eq:flow-lp}, and
\[
 r_{\pi'}(T)\le F^*,
 \qquad g^*\le F^*.
\]

It remains to bound the matches missed by $\pi$ when an inventory
is empty. Fix an edge $e=(i,j)$ with $f_e=f_{ij}>0$.
Let $A_e(t)$ count arrivals by time $t$ at the storage endpoint
for which the policy chooses edge $e$. Let $B_e(t)$ count arrivals
by time $t$ at the other endpoint for which the policy chooses
edge $e$.

Independently selecting each arrival of a rate-$a$ Poisson process
with a fixed probability $p$ produces a Poisson process of rate
$ap$. 
Here, supplier arrivals at $i$ are selected for edge $e$ with
probability $f_{ij}/\lambda_i$, and customer arrivals at $j$ with
probability $f_{ij}/\gamma_j$. The resulting rates are
\[
 \lambda_i\frac{f_{ij}}{\lambda_i}=
 \gamma_j\frac{f_{ij}}{\gamma_j}=f_e.
\]
Thus $A_e$ and $B_e$ are independent rate-$f_e$ Poisson processes.

Let $Q_e(t)$ be the number of agents in the inventory for $e$
at time $t$, after processing any arrival at that time.
Agents in this inventory have infinite patience and never abandon.
Each arrival counted by $B_e$ completes one match if the inventory
is nonempty immediately before arrival.

Let $\tau_{e,k}$ be the time of the $k$th arrival counted by $B_e$,
and define
\[
 L_e(t)
 :=\sum_{k=1}^{B_e(t)}
       \mathbf1_{\{Q_e(\tau_{e,k}-)=0\}}.
\]
Thus $L_e(t)$ counts arrivals in $B_e$ that find the inventory
empty immediately before arrival. These arrivals are never matched
under $\pi$. Every other arrival in $B_e$ completes one match, so
\[
 M_e^\pi(t)=B_e(t)-L_e(t).
\]
Starting empty, the inventory therefore satisfies
\[
 Q_e(t)=A_e(t)-M_e^\pi(t)
       =A_e(t)-B_e(t)+L_e(t).
\]

An arrival counted by $A_e$ increases the inventory by one.
An arrival counted by $B_e$ decreases it by one when it is positive;
otherwise, it increases $L_e$ by one and leaves the inventory empty.
Starting from zero, these updates show by induction over arrival
times that $L_e(t)$ equals the largest positive excess of $B_e$
over $A_e$ up to time $t$. In other words,
\begin{equation}
 L_e(T)=\sup_{0\le t\le T}(B_e(t)-A_e(t))^+.
 \label{pos:eq:reflection}
\end{equation}
Because $A_e$ and $B_e$ are independent Poisson processes with
the same rate, $B_e-A_e$ is a square-integrable martingale with
\[
 \E (B_e(T)-A_e(T))^2=2f_eT.
\]
The identity \eqref{pos:eq:reflection} and Doob's $L^2$ inequality
therefore give
\[
 \E  L_e(T)
 \le
 \left(\E \sup_{t\le T}|B_e(t)-A_e(t)|^2\right)^{1/2}
 \le 2\sqrt{2f_eT}.
\]
Using $M_e^\pi(T)=B_e(T)-L_e(T)$, we obtain
\[
 \E  M_e^\pi(T)
 =f_eT-\E  L_e(T)
 \ge f_eT-2\sqrt{2f_eT}.
\]
Multiplying by the nonnegative edge rewards and summing over
edges gives
\[
 \E  W^\pi(T)
 \ge TF^*-2\sqrt{2T}\sum_{e:f_e>0}w_e\sqrt{f_e}.
\]
Divide by $T$ and combine with the arrival upper bound to obtain
\eqref{pos:eq:rate-bound}. Both bounds converge to $F^*$,
so this single policy satisfies $g^\pi=g^*=F^*$, as claimed.

Finally, for rational input data, linear programming computes an optimal
flow in \eqref{eq:flow-lp} of polynomial encoding length in
polynomial time. Choosing storage endpoints and forming the
rational assignment probabilities therefore also take polynomial
time and give a polynomial-length policy description.
\end{proof}

\section{Proof of additional technical results} \label{app:technical-details}

\subsection{Proof of Lemma~\ref{dep:execution}}

\begin{proof}
We construct the required observation using discrete random choices.

\medskip
\noindent
\emph{1. Bound the population.}
Set
\[
 a:=\left\lceil\frac{4\Lambda T}{\rho}\right\rceil,
 \qquad h:=\Lambda+a.
\]
Stop the system if more than $a$ arrivals occur, and thereafter
report the empty occupied set. If $A(T)$ counts arrivals by
time $T$, then Markov's inequality gives
\[
 \Prob\{A(T)>a\}
 \le\frac{\Lambda T}{a}\le\frac{\rho}{4}.
\]
Thus stopping changes the observation law by total-variation
distance at most $\rho/4$. Before stopping, at most $a$ agents
are waiting, so the total arrival and abandonment rate is at
most $h$.

\medskip
\noindent
\emph{2. Sample a position in the event sequence.}
Represent the stopped system using the times of a rate-$h$
Poisson process. At each such time, choose each possible event
of current rate $r$ with probability $r/h$, and otherwise
make no change. Call the policy only after an actual arrival
or abandonment. Each event then has its required rate
$h(r/h)=r$; after stopping, make no further changes.

The event choices and policy decisions can be generated
independently of these times, since the policy uses only event
types, queue lengths, internal memory, and its own random bits.

Let $Z\sim\operatorname{Pois}(hT)$ count the generated times
in $[0,T]$. Conditional on $Z=z$, they are the ordered values
of $z$ independent uniform points in $[0,T]$.
The independent uniform observation $U_T$ therefore has each
of the $z+1$ positions among them with equal probability.
Consequently, the stopped-system observation can be sampled
by drawing $Z$, choosing $J$ uniformly from $\{0,\ldots,Z\}$,
and performing the first $J$ discrete steps.

To bound the number of steps, set
\[
 s:=\left\lceil\frac{4hT}{\rho}\right\rceil.
\]
Since
\[
 \Prob\{Z>s\}\le\frac{hT}{s}\le\frac{\rho}{4},
\]
we instead draw $\widehat Z$ from the conditional distribution
of $Z$ given $Z\le s$:
\begin{equation}\label{eq:sample-proba}
 \Prob\{\widehat Z=z\}
 =
 \frac{(hT)^z/z!}
      {\displaystyle\sum_{\ell=0}^{s}(hT)^\ell/\ell!},
 \qquad z=0,\ldots,s.
\end{equation}
Choose $J$ uniformly from $\{0,\ldots,\widehat Z\}$, perform
$J$ discrete steps, and return the resulting occupied supplier
set. Conditioning changes the observation law by
total-variation distance at most $\rho/4$.

\medskip
\noindent
\emph{3. Bound the computation.}
Since $\Lambda<2nL$, with $L$ and $\rho$ fixed, both $a$ and
$s$ are polynomial in the instance length and $T$.
The weights in \eqref{eq:sample-proba} can be computed recursively
and have polynomial encoding length. The event-selection
probabilities are also rational with polynomial encoding length.

A rational distribution can be sampled exactly in expected
polynomial time: express its probabilities with a common
integer denominator $D$, draw a uniform integer using
$\lceil\log_2D\rceil$ random bits, and repeat if it is at
least $D$. Each attempt succeeds with probability at least
$1/2$.

The simulation requires at most $s+2$ such samples and $s$
calls to the policy. Efficient executability therefore gives
a computable polynomial bound $R$ on its expected total
running time, including initialization.
Impose a limit of $\lceil4R/\rho\rceil$ operations and return
the empty set if it is exceeded. By Markov's inequality,
\[
 \Prob\{\text{time limit exceeded}\}
 \le\frac{R}{\lceil4R/\rho\rceil}\le\frac{\rho}{4}.
\]
The resulting algorithm has polynomial worst-case running
time. Its output $\widehat P$ satisfies
\[
 \bigl\|\mathcal L(\widehat P)-\mathcal L(P(U_T))\bigr\|_{\mathrm{TV}}
 \le
 \underbrace{\frac{\rho}{4}}_{\text{arrival cutoff}}
 +\underbrace{\frac{\rho}{4}}_{\text{conditioning}}
 +\underbrace{\frac{\rho}{4}}_{\text{time limit}}
 <\rho.
\]
\end{proof}

\subsection{Existence of a stationary distribution}  \label{comp:recurrence-section}

For completeness, we record the recurrence, moment, and convergence
properties of a stationary Markov policy used throughout the paper.

\begin{lemma}[Stationarity under unit abandonment]
\label{comp:recurrence}
Consider a finite-type matching system with independent Poisson
arrivals and a stationary Markov policy. Every agent who can wait
has independent rate-one exponential patience; all other agents have zero patience.
Let $Q(t)$ be the post-decision queue-length Markov chain,
and $N(q)$ the total population in state $q$.
Write $\lambda_{\rm w}$ for the total arrival rate of types that
can wait.

On the states reachable from empty, the chain is irreducible and positive recurrent, with a unique invariant law
$\nu$. Its law from empty converges to $\nu$ in total variation,
and
\begin{equation}
 \sup_{t\ge0}\E [N(Q(t))^2]
 \le\lambda_{\rm w}+\lambda_{\rm w}^2,
 \qquad
 \E_\nu N^2\le\lambda_{\rm w}+\lambda_{\rm w}^2.
 \label{comp:population-moments}
\end{equation}
Starting empty, the expectation of each queue length converges
to its stationary mean. 
\end{lemma}

\begin{proof}
On every bounded time interval there are finitely many
arrivals, and every abandonment or match removes previously
arrived agents. Hence the chain is nonexplosive.

From any state $q$, no arrivals occur during the next unit of
time and all initially waiting agents' patience expires with
probability
\[
 e^{-\Lambda}(1-e^{-1})^{N(q)}>0.
\]
On this event the system is empty after one unit of time.
Thus the states reachable from empty form a closed irreducible
class.

Arrivals increase the population at rate at most $\lambda_{\rm w}$,
abandonments decrease it at rate $N(q)$, and matching can only
decrease it further. Writing $\mathcal A$ for the generator of the chain, we have
\begin{equation}
 (\mathcal A N)(q)\le\lambda_{\rm w}-N(q).
 \label{comp:drift}
\end{equation}
The set $\{q:N(q)\le\lambda_{\rm w}+1\}$ is finite, and the
drift is at most $-1$ outside this set. The Foster--Lyapunov
criterion gives positive recurrence. The standard convergence
theorem for irreducible positive recurrent continuous-time
chains then gives a unique invariant law $\nu$ on the reachable
class and
\[
 \bigl\|\mathcal L(Q(t))-\nu\bigr\|_{\mathrm{TV}}
 \longrightarrow0.
\]

For the moment bounds, retain each waiting-type arrival in an
auxiliary system until its patience expires, including agents
already matched. Starting empty, its population $Y(t)$ satisfies
\[
 N(Q(t))\le Y(t),
 \qquad
 Y(t)\sim\operatorname{Pois}
 \bigl(\lambda_{\rm w}(1-e^{-t})\bigr).
\]
Consequently,
\[
 \sup_{t\ge0}\E [N(Q(t))^2]
 \le\sup_{t\ge0}\E[Y(t)^2]
 \le\lambda_{\rm w}+\lambda_{\rm w}^2.
\]
Total-variation convergence and monotone convergence give
\[
 \E_\nu[N^2]
 =\lim_{b\to\infty}\lim_{t\to\infty}
   \E [N(Q(t))^2\wedge b]
 \le\lambda_{\rm w}+\lambda_{\rm w}^2,
\]
proving \eqref{comp:population-moments}.

Finally, for any queue coordinate $Q_i$ and $b>0$, truncation
at $b$ yields
\begin{align*}
 \bigl|\E [Q_i(t)]-\E_\nu[Q_i]\bigr|
 &\le
 \bigl|\E [Q_i(t)\wedge b]-\E_\nu[Q_i\wedge b]\bigr|\\
 &\quad+\E [(Q_i(t)-b)^+]+\E_\nu[(Q_i-b)^+]\\
 &\le
 b\,\bigl\|\mathcal L(Q(t))-\nu\bigr\|_{\mathrm{TV}}
 +\frac{2(\lambda_{\rm w}+\lambda_{\rm w}^2)}{b}.
\end{align*}
Letting $t\to\infty$ and then $b\to\infty$ proves convergence
of each queue-length expectation to its stationary mean.
For the last inequality above, the identity
$x\wedge b=\int_0^b\mathbf1_{\{x>s\}}\,ds$ gives
\[
 \begin{aligned}
 &\bigl|\E[Q_i(t)\wedge b]-\E_\nu[Q_i\wedge b]\bigr|\\
 &\quad=
 \left|\int_0^b
 \bigl(\Prob\{Q_i(t)>s\}-\Prob_\nu\{Q_i>s\}\bigr)\,ds\right|\\
 &\quad\le
 \int_0^b
 \left|\Prob\{Q_i(t)>s\}-\Prob_\nu\{Q_i>s\}\right|\,ds\\
 &\quad\le
 b\,\bigl\|\mathcal L(Q(t))-\nu\bigr\|_{\mathrm{TV}}.
 \end{aligned}
\]
For the tail terms, use
$(x-b)^+\le x\mathbf1_{\{x>b\}}\le x^2/b$ for $x\ge0$; since each queue length is bounded by the total population,
the second-moment bounds give
\[
 \begin{aligned}
 &\E[(Q_i(t)-b)^+]+\E_\nu[(Q_i-b)^+]\\
 &\quad\le
 \frac{\E[Q_i(t)^2]+\E_\nu[Q_i^2]}{b}\\
 &\quad\le
 \frac{\E[N(Q(t))^2]+\E_\nu[N^2]}{b}\\
 &\quad\le
 \frac{2(\lambda_{\rm w}+\lambda_{\rm w}^2)}{b}.
 \end{aligned}
\]
\end{proof}

\section{The Finite-Zero regime}
\label{app:Finite-Zero}

We use the same cubic vertex-cover promise and incidence
construction as in the Finite-Finite proof. An edge customer
whose two compatible supplier queues are empty is now lost
immediately, which simplifies the upper bound. For the lower
bound, we again reserve one supplier at each vertex of a small
cover. A quadratic Lyapunov function bounds the loss of flexible
customers directly.

An initial version of this reduction was proposed with AI assistance
and subsequently simplified and revised by the authors; see the
AI methodology statement at the end of the paper.

\subsection{The construction}

Use $\varepsilon=1/100$ as in
Theorem~\ref{thm:cubic-source}. The source input is a finite
simple cubic graph $H=(V_H,E_H)$ and an integer $K$ satisfying
\begin{equation}
 n=|V_H|,\qquad m=|E_H|=3n/2,\qquad n/2\le K<n,
 \label{fz:source-range}
\end{equation}
under the promise that either $\tau(H)\le K$ or
$\tau(H)\ge(1+\varepsilon)K$.

For this section, fix
\begin{equation}
 \varepsilon:=\frac1{100},\qquad
 L:=\frac{96}{\varepsilon}=9600,\qquad
 n_0:=8\cdot10^{10}.
 \label{fz:constants}
\end{equation}
Replace $H$ by
\begin{equation}
 r:=\left\lceil\frac{n_0}{n}\right\rceil
 \label{fz:padding-factor}
\end{equation}
disjoint copies, and replace $K$ by $rK$. The copying factor is
bounded by a fixed constant, and the source promise is preserved.
After renaming the copied quantities, $n\ge n_0$. In particular,
\begin{equation}
 \frac{4L}{\varepsilon}\le\frac{\varepsilon n}{32},
 \label{fz:padding}
\end{equation}
because $128L/\varepsilon^2<n_0$.

For each vertex $i\in V_H$, create a supplier type $s_i$ with
arrival rate $L$ and independent rate-one exponential patience.
For each edge $e=\{i,j\}\in E_H$, create a customer type $c_e$
of arrival rate one, compatible with $s_i$ and $s_j$.
Finally, create a flexible customer type $c_0$, compatible with
every supplier, of arrival rate $\Gamma-m$, where
\begin{equation}
 a:=(1+\varepsilon/2)K,\qquad
 \Gamma:=nL-a.
 \label{fz:rates}
\end{equation}
All customers have zero patience, and every match earns one unit
of reward. The flexible arrival rate is positive since
$\Gamma-m>n(L-5/2-\varepsilon/2)>0$.

The compatibility graph is simple, connected, and bipartite.
It has $n$ supplier types, $m+1$ customer types, and
$2m+n=4n$ compatibility edges. Every supplier has degree four,
every edge customer has degree two, and the flexible customer
has degree $n$. The total arrival rate satisfies
\begin{equation}
 \Lambda=nL+\Gamma=2nL-a,\qquad
 nL<\Lambda<2nL.
 \label{fz:total-rate}
\end{equation}
All rates are rational and have polynomial encoding length,
and the construction takes polynomial time.

\subsection{The value gap and the hardness theorem}

\begin{theorem}[Finite-Zero hardness]
\label{thm:fz-hardness}
There exists a fixed rational constant $\eta>0$ such that the
following promise problem is NP-hard. Given a
Finite-Zero matching instance and a rational threshold $b>0$,
distinguish
\[
 \frac{g^*}{\Lambda}\ge b+\eta
 \qquad\text{from}\qquad
 \frac{g^*}{\Lambda}\le b-\eta.
\]
\end{theorem}

We use an upper bound valid for every admissible policy and
a lower bound attained by a policy constructed from a small
vertex cover.

\begin{proposition}[Finite-horizon upper bound]
\label{fz:upper}
For every admissible policy $\pi$ and every $T>0$,
\begin{equation}
 2\E M^\pi(T)\le[nL+\Gamma-\tau(H)]T.
 \label{fz:finite-upper}
\end{equation}
Consequently,
\begin{equation}
 g^*\le nL-\frac{a+\tau(H)}2.
 \label{fz:universal-value-bound}
\end{equation}
\end{proposition}

\begin{proposition}[Reward from a small cover]
\label{fz:lower}
If $\tau(H)\le K$, there is a deterministic stationary Markov
policy $\pi_U$ such that
\begin{equation}
 g^{\pi_U}\ge
 \Gamma-\frac{4m}{L}-\frac{\varepsilon n}{32}.
 \label{fz:lower-value}
\end{equation}
\end{proposition}

\begin{proof}[Proof of Theorem~\ref{thm:fz-hardness}]
Perform the construction above on the source input.
Define the two rational thresholds
\[
 g_-:=\Gamma-\frac{\varepsilon K}{4},
 \qquad
 g_+:=\Gamma-\frac{4m}{L}-\frac{\varepsilon n}{32}.
\]
Set
\[
 b_{H,K}:=\frac{g_++g_-}{2\Lambda},
 \qquad
 \eta:=\frac{\varepsilon}{256L}.
\]
Since $K\ge n/2$, $m=3n/2$, and $L=96/\varepsilon$,
\begin{align}
 g_+-g_-
 &=\frac{\varepsilon K}{4}
       -\frac{4m}{L}-\frac{\varepsilon n}{32}\notag\\
 &\ge\frac{\varepsilon n}{8}
       -\frac{\varepsilon n}{16}
       -\frac{\varepsilon n}{32}
 =\frac{\varepsilon n}{32}.
 \label{fz:separation}
\end{align}
By \eqref{fz:total-rate} and \eqref{fz:separation},
\[
 \frac{g_+-g_-}{2\Lambda}
 \ge\frac{\varepsilon n}{64\Lambda}
 >\frac{\varepsilon}{128L}>\eta.
\]

If $\tau(H)\le K$, Proposition~\ref{fz:lower} gives
\[
 \frac{g^*}{\Lambda}
 \ge\frac{g_+}{\Lambda}
 >b_{H,K}+\eta.
\]
If $\tau(H)\ge(1+\varepsilon)K$,
Proposition~\ref{fz:upper} gives
\begin{equation}
 \frac{g^*}{\Lambda}
 \le\frac{g_-}{\Lambda}
 <b_{H,K}-\eta.
 \label{fz:upper-value}
\end{equation}
Thus every produced instance satisfies the target promise.

Both the matching instance and the positive rational threshold
$b_{H,K}$ have polynomial encoding length and can be computed
in polynomial time from $(H,K)$. An algorithm distinguishing
the two target value ranges would therefore distinguish the
NP-hard source alternatives of Theorem~\ref{thm:cubic-source},
proving the claim.
\end{proof}

\subsection{A large cover forces lost reward}

\begin{proof}[Proof of Proposition~\ref{fz:upper}]
Write
\[
 P(t):=\{i\in V_H:Q_i^s(t)>0\},
 \qquad
 u(P):=|\{e\in E_H:e\cap P=\varnothing\}|.
\]
Lemma~\ref{lem:cover-completion} gives
\begin{equation}
 |P|+u(P)\ge\tau(H).
 \label{fz:cover-inequality}
\end{equation}
Supplier conservation and rate-one abandonment imply
\[
 \E M^\pi(T)
 \le nLT-\E \int_0^T\sum_iQ_i^s(t)\,dt
 \le nLT-\E \int_0^T|P(t)|\,dt.
\]

An edge customer arriving when both endpoints are empty
cannot be matched.
For each $e\in E_H$, let $N_e(t)$ count arrivals of
customers of type $c_e$. Then
\[
 \int_0^T
 \mathbf1_{\{e\cap P(t-)=\varnothing\}}\,dN_e(t)
\]
counts the arrivals of type $c_e$ by time $T$ that find both
compatible supplier queues empty.
By the Poisson compensation formula,
\[
 \E\int_0^T
 \mathbf1_{\{e\cap P(t-)=\varnothing\}}\,dN_e(t)
 =
 \E\int_0^T
 \mathbf1_{\{e\cap P(t-)=\varnothing\}}\,dt.
\]
Summing over $e\in E_H$, the expected number of these lost
customers is $\E\int_0^T u(P(t-))\,dt$. Since the expected total
number of customer arrivals is $\Gamma T$,
\[
 \E M^\pi(T)
 \le\Gamma T-\E \int_0^T u(P(t))\,dt.
\]
Here $P(t-)$ can be replaced by $P(t)$ because they differ
only at finitely many event times on $[0,T]$.

Adding the two inequalities and using
\eqref{fz:cover-inequality} proves \eqref{fz:finite-upper}.
Dividing by $2T$, taking the liminf, and then taking the
supremum over policies proves
\eqref{fz:universal-value-bound}.
\end{proof}

\subsection{A small cover serves nearly all demand}

Suppose $\tau(H)\le K$. Enlarge a cover, if necessary, to obtain
$U\subseteq V_H$ with $|U|=K$. For each edge $e\in E_H$, fix
an endpoint $o(e)\in e\cap U$, and define
\[
 \mathcal F_i:=\{e\in E_H:o(e)=i\},
 \qquad
 k_i:=|\mathcal F_i|.
\]
Then $k_i=0$ outside $U$, $0\le k_i\le3$, and $\sum_i k_i=m$.
The policy $\pi_U$ uses the following rules.
\begin{enumerate}
\item Each supplier arrival joins its queue.
\item A customer of type $c_e$ takes a supplier at $o(e)$
if that queue is nonempty, and otherwise leaves unmatched.
\item A flexible customer takes a supplier outside $U$, or
from a queue in $U$ containing at least two suppliers.
If neither is possible, it leaves unmatched.
\end{enumerate}
Use fixed queue orderings to resolve all choices.
The policy makes no voluntary discards and is a deterministic
stationary Markov policy. As in the Finite-Finite proof, define
\begin{equation}
 h_U(q^s):=\sum_{i\notin U}q_i^s
                 +\sum_{i\in U}(q_i^s-1)^+.
 \label{fz:eligible}
\end{equation}
This counts the suppliers eligible for flexible customers.

\begin{proof}[Proof of Proposition~\ref{fz:lower}]
By Lemma~\ref{comp:recurrence}, with $\lambda_{\rm w}=nL$,
the supplier queue process under $\pi_U$ has a unique invariant
law $\nu_U$ on the class reachable from empty, and its total
population has a finite stationary second moment.
Let $\mathcal A$ be its generator.
Its matching rate at state $q$ is
\[
 r_U(q):=(\Gamma-m)\ind{h_U(q^s)>0}
                  +\sum_i k_i\ind{q_i^s>0}
 \le\Gamma.
\]
Under $\pi_U$, every match occurs at a customer arrival.
An edge customer is matched exactly when its designated supplier
queue is nonempty, and a flexible customer is matched exactly
when eligible supplier inventory is positive. Thus the conditional
matching rate immediately before time $t$ is $r_U(Q(t-))$.
The compensation formula gives
\[
 \E M^{\pi_U}(T)
 =\E\int_0^T r_U(Q(t-))\,dt
 =\int_0^T\E r_U(Q(t))\,dt.
\]
The last equality follows from Fubini's theorem and the fact
that $Q(t-)$ and $Q(t)$ differ only at event times.
Moreover, $0\le r_U(q)\le\Gamma$, so total-variation convergence
to $\nu_U$ implies
\[
 \left|\E r_U(Q(t))-\E_{\nu_U}r_U(Q)\right|
 \le
 \Gamma\bigl\|\mathcal L_0(Q(t))-\nu_U\bigr\|_{\mathrm{TV}}
 \longrightarrow0.
\]
Taking time averages therefore yields
\begin{equation}
 r_{\pi_U}(T)
 =\frac1T\int_0^T\E r_U(Q(t))\,dt
 \longrightarrow\E_{\nu_U}r_U(Q)=g^{\pi_U}.
 \label{fz:stationary-reward}
\end{equation}

We now bound the two losses separately.
First consider edge customers. For $i\in U$, applying the
stationary generator identity to $\ind{q_i^s>0}$ gives
\[
 L\Prob_{\nu_U}(Q_i^s=0)
 =(1+k_i)\Prob_{\nu_U}(Q_i^s=1).
\]
Flexible customers cannot remove the last supplier at $i$,
so they do not contribute to this identity. Since
$\Prob_{\nu_U}(Q_i^s=1)\le1-\Prob_{\nu_U}(Q_i^s=0)$,
\begin{equation}
 \Prob_{\nu_U}(Q_i^s=0)
 \le\frac{1+k_i}{L+1+k_i}\le\frac4L.
 \label{fz:one-level}
\end{equation}
Consequently, the stationary loss rate of edge customers
is at most
\begin{equation}
 \sum_i k_i\Prob_{\nu_U}(Q_i^s=0)\le\frac{4m}{L}.
 \label{fz:edge-loss}
\end{equation}

It remains to bound the loss of flexible customers.
Let $N(q):=\sum_iq_i^s$ be the total supplier inventory,
and recall that $a=nL-\Gamma=(1+\varepsilon/2)K$.
Consider
\[
 V(q):=\bigl((a-N(q))^+\bigr)^2+N(q).
\]
At state $q$, total inventory increases at rate $nL$ and
decreases at rate $N(q)+r_U(q)$, with no downward transition
from zero. This statement uses the full queue process;
it does not require $N(Q(t))$ to be Markov.

For $v(s):=((a-s)^+)^2+s$ and $x:=(a-s)^+$,
\[
 v(s+1)-v(s)\le2-2x,
 \qquad
 v(s-1)-v(s)\le2x\quad(s\ge1).
\]
These inequalities follow because the derivative of
$((a-s)^+)^2$ is $-2(a-s)^+$ and is Lipschitz with constant
two. Taking $s=N(q)$ yields
\begin{align*}
 (\mathcal A V)(q)
 &\le nL(2-2x)+(N(q)+r_U(q))\,2x\\
 &=2nL-2x^2-2x\bigl(\Gamma-r_U(q)\bigr).
\end{align*}
By the reservation rule,
\begin{equation}
 h_U(q^s)=0\quad\Longrightarrow\quad N(q)\le K.
 \label{fz:stockout-implication}
\end{equation}
Hence $x\ge\varepsilon K/2$ whenever $h_U(q^s)=0$.
Moreover,
\[
 \Gamma-r_U(q)
 =(\Gamma-m)\ind{h_U(q^s)=0}
       +\sum_i k_i\ind{q_i^s=0}.
\]
All terms are nonnegative, so
\begin{equation}
 (\mathcal A V)(q)
 \le2nL-\varepsilon K(\Gamma-m)\ind{h_U(q^s)=0}.
 \label{fz:lyapunov-drift}
\end{equation}

The event rate is at most $\Lambda+N(q)$, and each jump
changes $V$ by at most a constant times $1+N(q)$.
The stationary second moment therefore makes the expected
sum of absolute generator terms finite.
Applying stationarity to bounded truncations of $V$ and
using dominated convergence justifies
$\E_{\nu_U}[\mathcal A V]=0$.
Taking expectations in \eqref{fz:lyapunov-drift} gives
\begin{equation}
 (\Gamma-m)\Prob_{\nu_U}(h_U(Q^s)=0)
 \le\frac{2nL}{\varepsilon K}
 \le\frac{4L}{\varepsilon}
 \le\frac{\varepsilon n}{32},
 \label{fz:flexible-loss-bound}
\end{equation}
where $K\ge n/2$ and \eqref{fz:padding} give the last two
inequalities.

The stationary matching rate equals $\Gamma$ minus the
loss rates of edge and flexible customers. Combining
\eqref{fz:stationary-reward} with the upper bounds in \eqref{fz:edge-loss} and
\eqref{fz:flexible-loss-bound} therefore proves the claim in \eqref{fz:lower-value}.
\end{proof}

\section*{AI methodology statement}

We used OpenAI's ChatGPT for research brainstorming, mathematical
discussions, proof development and checking, and manuscript drafting
and revision. In particular, an AI model proposed an initial
Vertex Cover reduction for the Finite-Zero regime. The authors
subsequently simplified and revised this construction substantially.
AI assistance also contributed to the presentation and refinement of 
the remaining proofs. 

The authors reviewed and revised the resulting material and take
full responsibility for the correctness, originality, and integrity
of the manuscript, including all content developed with AI assistance.

\clearpage
\bibliographystyle{matchingalpha}
\bibliography{reference}
\end{document}